\documentclass{aa}  

\usepackage{graphicx}
\usepackage{xcolor}
\usepackage{txfonts}
\usepackage{tikz}
\usetikzlibrary{matrix}

\newcommand{\abs}[1]{\left\lvert#1\right\rvert}
\newcommand{\ip}[1]{\left\langle #1 \right\rangle}
\begin{document} 
\def\d{\displaystyle}
\def\n{\noindent}	
\def\t{\tau}
\def\p{\varphi}
\def\k{\kappa}
\def\pa{\partial}
\def\o{\omega}
\def\D{\Delta}
\def\l{\left}
\def\r{\right}
\def\c{\nabla \cdot \pmb{\xi}}
\def\e{\exp\l\{i \left( k_y y + k_z z \right) \r\}}
\def\ov{\overline}

   \title{Stability of solar atmospheric structures harboring standing slow waves}
   
   \subtitle{An analytical model in a compressible plasma}

   \author{M. Geeraerts \inst{1}
          \and
          T. Van Doorsselaere \inst{1}
          }

   \institute{Centre for mathematical Plasma Astrophysics (CmPA), Mathematics Department, KU Leuven, Celestijnenlaan 200B bus 2400, B-3001 Leuven, Belgium
             }

   \date{}

% \abstract{}{}{}{}{} 
% 5 {} token are mandatory
 
  \abstract
  % leave it empty if necessary  
   {In the context of the solar coronal heating problem, one possible explanation for the high coronal temperature is the release of energy by magnetohydrodynamic (MHD) waves. The energy transfer is believed to be possible, among others, by the development of the Kelvin-Helmholtz instability (KHI) in coronal loops.}
   {Our aim is to determine if standing slow waves in solar atmospheric structures such as coronal loops, and also prominence threads, sunspots, and pores, can trigger the KHI due to the oscillating shear flow at the structure's boundary.}
   {We used linearized nonstationary MHD to work out an analytical model in a cartesian reference frame. The model describes a compressible plasma near a discontinuous interface separating two regions of homogeneous plasma, each harboring an oscillating velocity field with a constant amplitude which is parallel to the background magnetic field and aligned with the interface. The obtained analytical results were then used to determine the stability of said interface, both in coronal and photospheric conditions.}
   {We find that the stability of the interface is determined by a Mathieu equation. In function of the parameters of this equation, the interface can either be stable or unstable. For coronal as well as photospheric conditions, we find that the interface is stable with respect to the KHI. Theoretically, it can, however, be unstable with respect to a parametric resonance instability, although it seems physically unlikely. We conclude that, in this simplified setup, a standing slow wave does not trigger the KHI without the involvement of additional physical processes.}
   {}


   \maketitle
%
%-------------------------------------------------------------------

\section{Introduction}

Although it has been known for a long time that the corona is three orders of magnitude hotter than the photosphere, a definite explanation for this unexpected feature has continued to elude researchers. One possible explanation that has been advanced over the years, and which is supported by both observations and theoretical models, is the deposition of energy of magnetohydrodynamic (MHD) waves into the corona \citep{Parnell&DeMoortel2012, Arregui2015, VanDoorsselaereEtAl2020}. There are several mechanisms that allow for the conversion of wave energy into heating of the surrounding plasma. Possibilities include dissipation by Ohmic resistivity, values of which were calculated by \citet{Kovitya&Cram1983} for the photosphere and used by \citet{ChenEtAl2018}, \citet{GeeraertsEtAl2020}, and \citet{ChenEtAl2020} to infer its importance in damping slow waves in a photospheric pore. Mode coupling \citep{PascoeEtAl2010, PascoeEtAl2012, HollwegEtAl2013, DeMoortelEtAl2016} and resonant absorption, both in the Alfv\'en \citep{Hollweg&Yang1988, HollwegEtAl1990, GoossensEtAl1992, GoossensEtAl2002a, SolerEtAl2013} and cusp \citep{CadezEtAl1997, ErdelyiEtAl2001,SolerEtAl2009c, YuEtAl2017b} continua, have also been studied for their potential to damp waves by transferring their energy to local oscillations. Another mechanism for damping waves in atmospheric structures and which has received considerable attention recently is the Kelvin-Helmholtz instability (KHI). Indeed, the transition of wave energy to turbulence and plasma heating on smaller scales in the coronal plasma is known to be facilitated by this instability \citep{Heyvaerts&Priest1983, OfmanEtAl1994, KarpenEtAl1994, KarampelasEtAl2017, AfanasyevEtAl2019, HillierEtAl2020, ShiEtAl2021}. The KHI has been observed in the solar atmosphere, for example on coronal mass ejection flanks \citep{Ofman&Thompson2011, FoullonEtAl2011}, and, more recently, on coronal loops \citep{SamantaEtAl2019}.

Slow magnetosonic waves were observed in  solar coronal loops more than a decade ago, both as propagating waves and standing waves. Upwardly propagating disturbances along coronal loops have been observed, for example, by \citet{Berghmans&Clette1999}, \citet{NightingaleEtAl1999}, and \citet{DeMoortelEtAl2000}. These have been interpreted as propagating slow modes through a theoretical model by \citet{NakariakovEtAl2000}. Other observations of disturbances in coronal loops by the SUMER spectrometer have been analyzed and interpreted by \citet{WangEtAl2002, WangEtAl2003a, WangEtAl2003b, WangEtAl2007} as standing slow modes, whereas \citet{KumarEtAl2013} and \citet{MandalEtAl2016} reported the observation of a reflecting, also referred to as sloshing, slow wave in a coronal loop. \citet{Wang2011} provides a review of observations and modeling of standing slow waves in coronal loops. These modes have also been studied through numerical simulations, for example, by \citet{DeMoortel&Hood2003}, who found that thermal conduction is an important damping mechanism for propagating slow waves in coronal conditions, and by \citet{MandalEtAl2016}, who reported about the frequency dependent damping of propagating slow waves in a coronal loop. Structures in the lower solar atmosphere, such as sunspots and pores in the photosphere, have also been observed to harbor slow magnetosonic waves \citep{DorotovicEtAl2008, DorotovicEtAl2014, MortonEtAl2011, GrantEtAl2015, MoreelsEtAl2015, FreijEtAl2016}, which were then classified as either surface or body waves \citep{MoreelsEtAl2013, KeysEtAl2018}.

The KHI and its growth rate at the interface between two aligned sheared stationary flows have been known for a long time \citep{Chandrasekhar1961}. Although in the purely hydrodynamics (HD) model the instability always develops in the presence of a shear flow, in the MHD model an aligned magnetic field can prevent its triggering. A natural question that then arises in the context of wave heating is under which conditions the KHI would develop in the presence of an oscillating shear flow. Indeed, it is possible that waves propagating or standing in solar coronal structures, such as loops, can trigger the KHI on the boundary and hereby convey some of their energy to the plasma background. \citet{ZaqarashviliEtAl2015}, for example, studied the KHI in rotating and twisted magnetized jets in the solar atmosphere in the presence of a stationary shear flow and used their derivations to discuss the stability of standing kink and torsional Alfv\'en waves at the velocity antinodes. They found that the standing waves are always unstable whereas the propagating waves are stable. Transverse oscillations are known to be unstable to the KHI in coronal loops and numerical studies on this topic include \citet{TerradasEtAl2008}, \citet{AntolinEtAl2014, AntolinEtAl2015, AntolinEtAl2017}, \citet{MagyarEtAl2015}, \citet{GuoEtAl2019}, \citet{KarampelasEtAl2019} and \citet{PascoeEtAl2020}.

There have also been several analytical studies regarding the stability of the interface between oscillating sheared flows. \citet{Kelly1965} looked at the HD case of two parallel sheared flows aligned with the interface, whereas \citet{Roberts1973} studied the same setup in the MHD model. More recently, \citet{BarbulescuEtAl2019} and \citet{HillierEtAl2019} investigated the stability of transverse oscillations in coronal loops by modeling them locally at the loop boundary as a cartesian interface between sheared background flows, with the background velocity perpendicular to the background magnetic field. Each of these studies revealed that the interface between oscillating sheared flows is always unstable, in contrast to the constant sheared flows case. All of these studies relying on the simplifying assumption that the fluid is incompressible, it is worth asking whether their conclusions remain unchanged when compression is included.

The goal of this paper is therefore twofold. Firstly, it aims at extending the known incompressible model of \citet{Roberts1973} for a plasma with an oscillating velocity field aligned with both the magnetic field and the interface to a compressible version. The focus lies on finding expressions for the eigenfunctions and, in particular, to derive their evolution over time. The interest in doing this lies in identifying the shortcomings made by the approximation of the incompressible model, at the cost of considerably more involved analytical derivations. This is the subject of Sections 2 and 3. Secondly, it aims at expressing more general instability conditions as compared to the incompressible model, by taking into account the subtleties arising due to the inclusion of compression. In particular, it will allow us to assess the stability of certain solar atmospheric structures harboring a standing slow wave. We will do so by using this model as a local approximation of the structure's boundary, at the position where the velocity shear is the greatest (for instance in a cusp resonance). We will also compare our findings for the local stability of slow waves to those of \citet{BarbulescuEtAl2019} and \citet{HillierEtAl2019} for the local stability of fast kink waves. This is the subject of Section 4.

\section{Model}\label{Model}

We derived an analytical model for a compressible plasma at the boundary of solar atmospheric structures which can, in a first rough approximation, be modeled as a cylinder with a discontinuous boundary separating two regions of homogeneous plasma with different properties. Such structures include coronal loops, prominence threads, sunspots and photospheric pores. 
%The background magnetic field, assumed straight and aligned with the loop's axis, is parallel to the interface. Furthermore, we included a parallel oscillating velocity on both sides of the interface and with different constant amplitudes to model the region at an antinode of the perturbed velocity of a standing slow wave in the loop. We then determined the stability of the interface.%
The physical setup here is the same as in \citet{Roberts1973}, except we included compression. We point out that, in a more realistic setup, a smooth transition layer would have to be included at the interface. This would result in the possibility of resonance occuring between the main oscillation of the structure and local slow mode oscillations in the boundary layer, as studied theoretically by \citet{YuEtAl2017b}. The analytical derivations of \citet{GoossensEtAl2021} show that, when slow waves are resonantly absorbed in the cusp continuum, both the azimuthal component of vorticity and the parallel component of the plasma displacement are large. The huge amount of vorticity could indicate the possibility of the KHI developing in those conditions. However, the sharp spatial variation of the parallel displacement and the truly discontinuous interface separating two plasma regions are absent from the present model. This should be kept in mind when drawing conclusions regarding shear flows in resonances.

\begin{figure}
   \centering
   \includegraphics[scale=0.155]{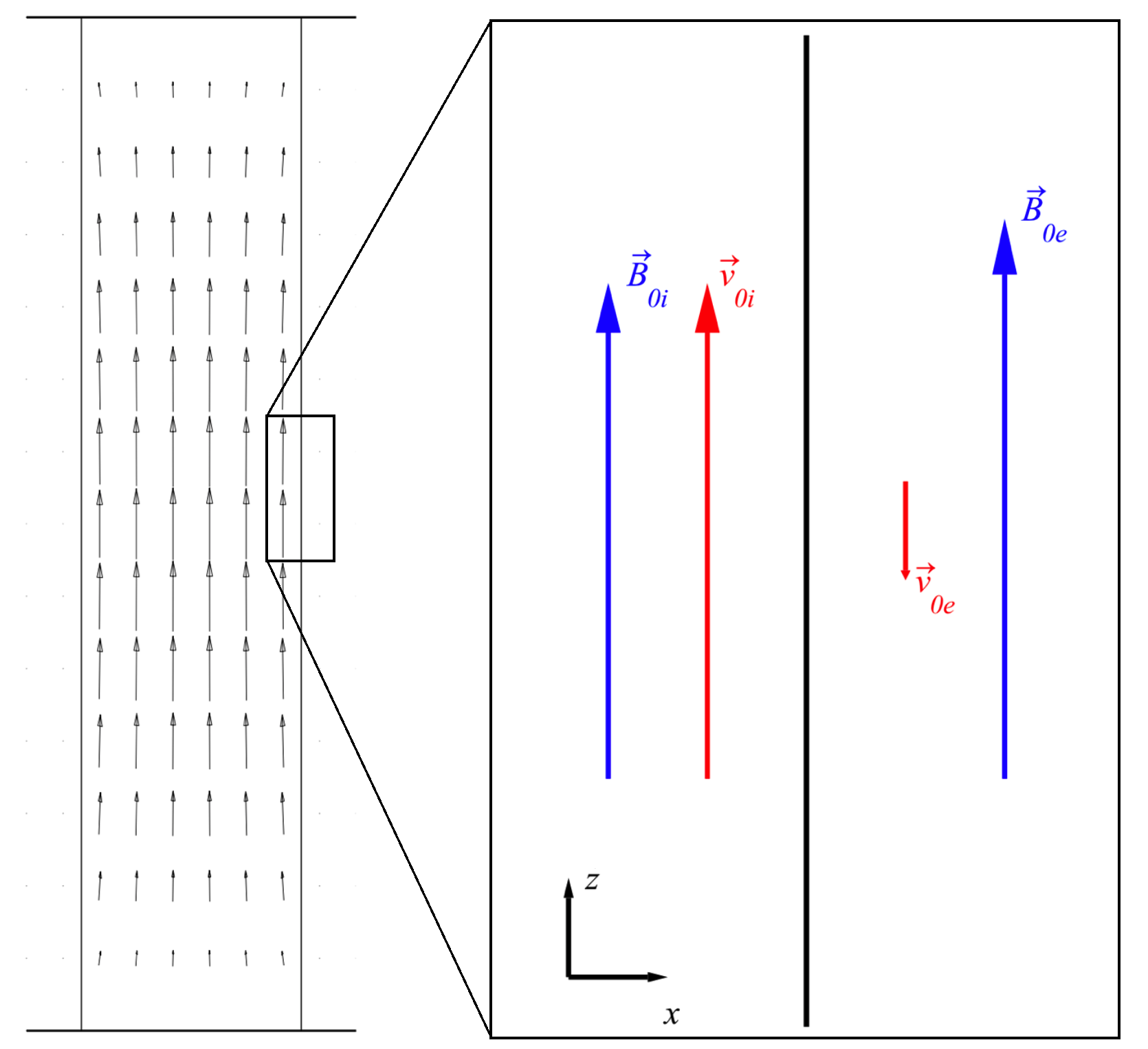}
      \caption{Sketch of a longitudinal cut along the axis of a coronal loop harboring a fundamental standing slow body sausage mode, at the velocity's maximal amplitude (left) and the local cartesian model at the boundary (right). The arrows on the left represent the velocity field and their lengths are to scale with the relative local magnitude of the field for a slow body sausage mode at a given time. The lengths and directions of the magnetic field arrows on the right figure are consistent for that same slow mode, whereas for the velocity arrows only the direction is consistent (the length of the exterior velocity arrow having been increased for visual clarification).}
         \label{SBSM}
   \end{figure}

In order to be able to make progress in the analytical derivations, the model uses a Cartesian coordinate system $(x, y, z)$, where $x=0$ is the interface and the $z$-direction is the longitudinal direction (i.e., along the cylinder's axis). The region $x<0$ represents the interior of the structure, whereas the region $x>0$ represents the surrounding plasma. This is a model for the local stability at the boundary, in the region of the structure where the shear in longitudinal velocity is the greatest, that is to say, at an antinode of the longitudinal component of the velocity eigenfunction. For the longitudinally fundamental slow mode, this would be at the middle of the structure (see Figure \ref{SBSM}), whereas for the first longitudinal overtone, for example, this would be at a quarter of the structure's length measured from either end. The time variable is denoted by $t$.

After linearizing the ideal MHD equations, the equilibrium quantities are denoted by the subscript $0$ and the perturbed quantities are denoted by the subscript $1$. The regions on each side of the interface are two homogeneous but different plasmas. This means each region has its own values for the background quantities, which are assumed spatially constant. The background magnetic field is assumed to be a straight and constant axial field along the $z$-coordinate: $\pmb{B}_0 = B_{0z} \pmb{1}_z$. Furthermore, we assume that the background flow is oscillating with a certain frequency $\omega_0$, which represents the frequency of the standing slow wave: $\pmb{v}_0 = V_0 \cos \left( \omega_0 t \right) \pmb{1}_z$. It would, of course, be more accurate to also include a background oscillation for the other quantities, such as magnetic field, pressure, and density. In the context of solar atmospheric structures, the magnetic field oscillations that occur because of this external forcing in the background can, however, be neglected in a first approximation with respect to the strong longitudinal magnetic field. As for the density and pressure background oscillations, they can be neglected in this model because they are in antiphase with respect to the velocity in their longitudinal profile and thus have a node where the logitudinal component of velocity has an antinode. For slow modes, the longitudinal component of the velocity is typically much larger than its other components (see left of Figure \ref{SBSM}), which can thus be neglected at the former's antinode as well.

The perturbed density, thermal pressure, velocity and magnetic field are denoted with $\rho_1$, $p_1$, $\pmb{v}_1$ and $\pmb{B}_1$, respectively. In what follows the perturbed eulerian total pressure will be used as well, which is given by $P_1 = p_1 + \frac{\pmb{B}_0 \cdot \pmb{B}_1}{\mu_0}$. Although gravity certainly has a role in solar atmospheric wave dynamics, it is neglected in this model in order to make some analytical progress.

The linearized compressible ideal MHD equations can, under these assumptions, be written as follows:

\begin{align}
&\d\frac{D \rho_1}{D t} +\rho_0 \l( \nabla \cdot \pmb{v}_1 \r) = 0\text{,} \label{eq1}\\
& \rho_1 \frac{\partial \pmb{v}_0}{\partial t}  + \rho_0 \frac{D \pmb{v}_1}{D t} = -\nabla P_1 + \frac{1}{\mu_0} \left( \pmb{B}_0 \cdot \nabla \right) \pmb{B}_1 \text{,} \label{eq2}\\
&\frac{D \pmb{B}_1}{D t} = - \pmb{B}_0 \left( \nabla \cdot \pmb{v}_1 \right) + \left( \pmb{B}_0 \cdot \nabla \right) \pmb{v}_1 \text{,} \label{eq3}\\
&\frac{D p_1}{D t} + \rho_0 v_{\text{s}}^2 \l( \nabla \cdot \pmb{v}_1 \r) = 0\text{,} \label{eq4}
\end{align}
where $\frac{D f}{Dt} = \frac{\partial f}{\partial t} + \left(\pmb{v}_0 \cdot \nabla\right) f$ is the Lagrangian derivative of a quantity $f$, $v_{\text{s}} = \sqrt{\frac{\gamma p_0}{\rho_0}}$ is the speed of sound and $\mu_0$ the magnetic permeability of free space. The first equation, Eq. \eqref{eq1}, has this simpler form because $\nabla \rho_0 = 0$ in each region (i.e., both inside and outside the cylinder). Since the background quantities depend only on $x$ and $t$, the perturbed quantities can be Fourier-analyzed in the $y$ and $z$ coordinates and are thus assumed to have the following form: $f_1 = \overline{f}_1(x,t) \e$, for each of the perturbed quantities $p_1$, $\rho_1$, $v_{1x}$, $v_{1 y}$, $v_{1z}$, $B_{1x}$, $B_{1 y}$ and $B_{1z}$. 

\section{Expressions for the perturbed quantities}

In this section, we derive the governing equations for the evolution of linear MHD perturbations in a compressible plasma for the model described in the previous section. In the next section, we then try to use the obtained information to describe the stability of the interface in compressible plasma conditions.

\subsection{The central quantity $\c$}

In what follows, an expression is derived for the compression term $\c$. This term is central to finding expressions for all other physical quantities, in the sense that they can all be derived solely from it.

By using the Lagrangian displacement $\pmb{\xi}$ defined by $\frac{D \pmb{\xi}}{D t} = \pmb{v}_1$, it can be shown (see Appendix \ref{A1}) that the compression term $\c$ must satisfy the following partial differential equation:

\begin{align}
&\d\frac{D^4 \l( \c \r)}{Dt^4} \; + \; i k_z \o_0 V_0 \frac{D^2 \l( \sin \l(\o_0 t \r)  \l( \c \r) \r)}{Dt^2} \notag\\
& \qquad \; - \; \l( v_A^2 + v_s^2 \r) \frac{D^2}{Dt^2} \l( \frac{\pa^2 \l( \c \r)}{\pa x^2} \r) \; + \; k^2 \l( v_A^2 + v_s^2 \r) \frac{D^2 \l( \c \r)}{Dt^2} \notag\\
& \qquad \; - \; i k_z V_0 \o_0 \sin \l( \o_0 t \r) v_A^2 \frac{\pa^2 \l( \c \r)}{\pa x^2} \notag\\
& \qquad \; + \; i k_z  k^2 V_0 \o_0 \sin \l(\o_0 t \r)  v_A^2 \l( \c \r) \; - \; k_z^2 v_A^2 v_s^2 \frac{\pa^2 \l( \c \r)}{\pa x^2} \notag \\
& \qquad \; + \; k_z^2 k^2 v_A^2 v_s^2 \l( \c \r) = 0 \text{,} \label{Eqdivxi}
\end{align}
where $k = \sqrt{k_y^2 + k_z^2}$ and $v_A = B_{0z} / \sqrt{\mu_0 \rho_0}$ is the Alfv\'en speed. This partial differential equation is of fourth order in $t$ and second order in $x$. Note that if $V_0 = 0$, such that the time dependence of the background quantities disappears, we retrieve an ordinary differential equation (ODE) of second order in $x$ which is the governing equation for the compression term $\c$ in a plasma without a background flow. This is the equation used for example in \citet{Edwin&Roberts1983} to study wave modes in a magnetic cylinder in the framework of ideal MHD in a plasma without background flow.

By drawing inspiration from the derivations of \citet{BarbulescuEtAl2019}, one finds that Eq. \eqref{Eqdivxi} can take a simpler form if $\c$ is written as

\begin{equation}\label{f-g}
\c \; = \; f(x,t) g(t) \e \text{,}
\end{equation}
where $g(t) = \exp \l\{ \frac{-i k_z V_0 \sin \l( \o_0 t \r)}{\o_0} \r\}$. Since $g(t)$ has modulus equal to $1$ for all $t \in \mathbb{R}$, the stability of $\c$ over time is entirely determined by $f$. Inserting expression \eqref{f-g} into Eq. \eqref{Eqdivxi} simplifies this equation considerably, now taking the following form:

\begin{align}
& {\frac {\partial ^{4} f(x,t)}{\partial {t}^{4}}} \; - \; \left( {v_{A}}^{2}+{v_{s}}^{2} \right) {\frac {\partial ^{4} f(x,t)}{
\partial {x}^{2}\partial {t}^{2}}}  \notag\\
& \qquad \; + \; \left[ i \omega_{0} k_{z} V_{0} \sin \left( \omega_{0} t \right) + k^2 \left( {v_{A}}^{2}+{v_{s}}^{2} \right) \right] {\frac {\partial ^{2} f(x,t)}{\partial {t}^{2}}} \notag \\
 & \qquad \; - \; \left[ i \omega_{0} k_{z} V_{0} {v_{A}}^{2} \sin \left( \omega_{0} t \right)+{k_{z}}^{2}{v_{A}}^{2}{v_{s}}^{2}
 \right] {\frac {\partial ^{2} f(x,t)}{\partial {x}^{2}}} \notag \\
 & \qquad \; + \; 2 i \omega_{0}^2 k_{z} V_{0} \cos \left( \omega_{0} t \right) {\frac {\partial f(x,t)}{\partial t}} \notag \\
 & \qquad \; + \; \l[ i \omega_0 k_z V_0 \sin \l( \omega_0 t \r) \l( k^2 v_A^2 - \o_0^2 \r) \r. \notag\\
 & \l. \hspace{4.5cm} + k^2 {k_{z}}^{2}{v_{A}}^{2}{v_{s}}^2 \r] f(x,t) = 0 \text{.} \label{Eqf(x,t)}
\end{align}
Equation \eqref{Eqf(x,t)} has a solution in the form of $F(x)G(t)$, where $F$ and $G$ satisfy the following ODEs (with $m$ a constant):

\begin{equation} \label{EqF}
\d\frac{d^2 F(x)}{dx^2} \; + \; m^2 F(x) \; = \; 0
\end{equation}
and

\begin{align}
&\d\frac{d^4 G(\t)}{d \t^4} \; + \; \l( a_1 + q_1 \sin \l( \t \r) \r) \frac{d^2 G(\t)}{d \t^2} \; + \; q_3 \cos \l( \t \r)  \frac{d G(\t)}{d \t} \notag \\
& \hspace{4.0cm} + \; \l( a_2 + q_2 \sin \l( \t \r) \r) G(\t) \; = \; 0 \text{,} \label{EqG}
\end{align}
with $\t = \o_0 t$, and where

\begin{align*}
&\qquad a_1 = \frac{\l(v_A^2 + v_s^2 \r) \l(m^2 + k^2 \r)}{\o_0^2} \text{, } \;\;  q_1 = \frac{i k_z V_0}{\o_0} \text{, } \\
&\qquad q_3 = \d\frac{2 i k_z V_0}{\o_0} \text{,} \\
&\qquad a_2 = \frac{k_z^2 v_A^2 v_s^2 \l( m^2 + k^2 \r)}{\o_0^4} \text{, } \;\;\;\;\;  q_2 = \frac{i k_z V_0 \l[ \l( m^2 + k^2 \r) v_A^2 - \o_0^2 \r] }{\o_0^3} \text{.}
\end{align*}
These two equations are related by the constant $m$, which occurs in both of them.

\subsubsection{The spatial function $F$}

Equation \eqref{EqF} has a simple analytical solution, namely

\begin{equation}\label{SolF}
F(x) \; = \; C_1 \exp \l( i m x \r) \; + \; C_2 \exp \l( -i m x \r) \text{,}
\end{equation}
with $C_1$ and $C_2$ arbitrary constants. From this it can be inferred that $m$ plays the role of the wavenumber along $x$, which is the direction normal to the interface. The focus of this paper goes to standing slow waves in solar atmospheric structures that can be modeled as a straight cylinder with a circular base and a discontinuous boundary separating two homogeneous but different plasma regions. Therefore, each region has its own value for $m$ (namely $m_i$ for the interior and $m_e$ for the exterior), each being able to only take on specific values depending on the boundary conditions at the interface. Finding expressions for $m_i$ and $m_e$ is, however, not straightforward. This will be discussed in Section \ref{Gfunction}. We note that for the purpose of studying the local stability of the interface, $m$ will be taken purely imaginary in order to have evanescent spatial profiles in the normal direction.

\subsubsection{The temporal function $G$}\label{Gfunction}

Equation \eqref{EqG} does not have a simple analytical solution. This equation governs the stability of the compression term $\c$ over time, its parameters depending on the background quantities $v_A$, $v_s$, $V_0$ and $\o_0$. The function $h(t) = G(t) g(t)$, where $G$ is the solution to equation \eqref{EqG}, actually describes the general time evolution of $\c$ in a compressible homogeneous plasma of inifinite extent with an oscillating background velocity which is parallel to the straight and constant equilibrium magnetic field. Although there is no closed-form analytical solution, the boundedness of $\c$ over time can be derived from the properties of Eq. \eqref{EqG}. Indeed, this linear fourth-order ODE has periodic coefficients with the same period and hence it falls in the category of equations which obey Floquet theory \citep{Chicone2008}. 

Defining $\tau = \o_0 t$, it can be rewritten as a $4 \times 4$ system of first-order ODEs of the form $\pmb{x}'(\t) = A(\t) \pmb{x}(\t)$, where the coefficient matrix $A(\t)$ is periodic with a period of $2 \pi$. One can find four linearly independent fundamental solution vectors $\pmb{x}_1(\t)$, $\pmb{x}_2(\t)$, $\pmb{x}_3(\t)$ and $\pmb{x}_4(\t)$, which depend on their respective initial conditions. The matrix obtained by putting these four vectors into its columns is called a fundamental solution matrix. If we take as initial condition matrix the identity matrix, Floquet theory states that the corresponding fundamental solution matrix (which we denote by $X(\t)$) evaluated at one period (i.e., at $\tau = 2 \pi$ in our case) is intimately linked to the boundedness of the solutions of the ODE. Indeed, denoting the eigenvalues of $X(2\pi)$ by $\lambda_1$, $\lambda_2$, $\lambda_3$ and $\lambda_4$, Floquet's theorem states that for each distinct eigenvalue $\lambda$, if we write $\lambda = e^{2 \pi \mu}$ where $\mu \in \mathbb{C}$ is called the Floquet exponent, there exists an independent solution of the system which has the form

\begin{equation} \label{FloquetSol}
\pmb{x}(\t) = e^{\mu \t} \pmb{\Phi}(\t) \text{,}
\end{equation}
where $\pmb{\Phi}$ contains $2 \pi$-periodic functions. This means that a solution to Eq. \eqref{EqG} is unstable if and only if $\abs{\lambda} > 1$ for at least one eigenvalue $\lambda$ of $X(2 \pi)$. If there are less than four distinct eigenvalues of $X(2 \pi)$, it is possible there are less than four independent Floquet solutions of the form \eqref{FloquetSol}. If the eigenspace of $X(2 \pi)$ has dimension $4$, there are still four independent eigenvectors $\pmb{\Phi}_i$ and thus four independent Floquet solutions. If the eigenspace of $X(2 \pi)$ has dimension less than $4$, then there are less than four independent Floquet solutions. In that case, extra independent solutions have to be found by using the Jordan normal form of $X(2 \pi)$ (see e.g., \citet{Cesari1971} for more details). These extra independent Jordan solutions are always unstable. The stability of $\c$ is thus entirely determined by the Floquet exponents $\mu$.

A solution to Eq. \eqref{EqG} can be written in the form of a series. Knowing that each independent Floquet solution has the form $e^{\mu \t} \Phi(\t)$ with $\Phi$ a $2 \pi$-periodic function, one can assume that such a basic independent solution can be written as

\begin{equation}\label{Gseries}
G(\t) \; = \; e^{\mu \t} \d\sum_{j = -\infty}^{\infty} \varphi_j \; e^{i j \t} \text{,}
\end{equation} 
where the $\varphi_j$ are unknown coefficients. Writing $\mu=i \nu$ for convenience, the following recurence relation between the coefficients $\varphi_j$ can then be derived by inserting expression \eqref{Gseries} into Eq. \eqref{EqG}:

\begin{equation} \label{recursion}
- \varepsilon_j \; \varphi_{j-1} \;\; + \;\; \varphi_j \;\; + \;\; \varepsilon_j \; \varphi_{j+1} \;\; = \;\; 0\text{,} 
\end{equation}
where

\begin{equation} \label{epsj}
\varepsilon_j \; = \; \d\frac{\frac{1}{2} k_z V_0 \o_0 \l[ \l( j + \nu \r)^2 \o_0^2 - K^2 v_A^2 \r]}{\l( j + \nu \r)^4 \o_0^4 - \l( j + \nu \r)^2 \o_0^2 K^2 \l( v_A^2 + v_s^2 \r) + k_z^2 K^2 v_A^2 v_s^2}
\end{equation}
with $K = \sqrt{k^2 + m^2}$. Equations \eqref{recursion} are nontrivially satisfied if and only if the following infinite determinant vanishes:

\begin{equation} \label{Delta}
\Delta = \begin{vmatrix}
\ddots & \ddots & &  &  0\\ \ddots & 1 & \varepsilon_{-1} &  & \\  & -\varepsilon_0 & 1 & \varepsilon_0 & \\ &  & -\varepsilon_1 & 1 & \ddots \\ 0 &  &  & \ddots & \ddots
\end{vmatrix} \text{.}
\end{equation}
This is a Fredholm determinant. Denoting with $A$ the operator defined by the corresponding infinite matrix, this Fredholm determinant is well-defined if the operator $A-I$ (where $I$ is the identity operator) is a trace class operator on $\ell^2(\mathbb{Z})$, the Hilbert space of square-summable sequences of complex numbers with entire index. It can be shown that this is the case here (see for example \citet{Strang2005} for the method to follow), except if the denominator of one of the $\varepsilon_j$ vanishes. This happens if the perturbation, which is a normal mode with wave vector $\pmb{K} = (m, k_y, k_z)$, satisfies the equation

\begin{equation} \label{res}
(j+\nu)^2 \o_0^2 \; = \; \d\frac{K^2 \l(v_A^2 + v_s^2 \r)}{2} \l\{ 1 \pm \l[ 1 - \d\frac{4 k_z^2 v_A^2 v_s^2}{K^2 \l( v_A^2 + v_s^2 \r)^2} \r]^{1/2}  \r\} \text{,}
\end{equation}
for a $j \in \mathbb{Z}$. This represents a resonance between the background oscillator with frequency $\o_0$ and a magnetosonic wave, in a homogeneous plasma of infinite extent. In this case $m$ is a free parameter (like $k_y$ and $k_z$) and can potentially take any real value. In the model we describe in this paper, with an interface separating two such homogeneous plasmas, only certain specific values of $m$ (different in both regions) are physically possible. Because surface waves on the interface are actually of interest in this case, $m$ should be imaginary. In Section \ref{StabInterf} we study the resonance of these surface waves with the background oscillator.

 Considering $\Delta$ as a function of the two unknowns $\nu$ and $m$, it is thus well-defined for every $(\nu,m)$ except in the poles of the $\varepsilon_j$. It can also easily be checked with Eqs. \eqref{epsj} and \eqref{Delta} that $\Delta(\nu, m) = \Delta(\nu +1, m)$ and $\Delta(\nu, m) = \Delta(- \nu, m)$. As previously stated, there are in general four independent Floquet solutions of the form \eqref{Gseries} to Eq. \eqref{EqG}. Hence, all solutions for $\nu$ of $\Delta(\nu,m) = 0$ (as functions of $m$) which correspond to distinct solutions of the differential equation lie on the strip $-0.5 < \text{Re}\l[ \nu \r] \leq 0.5$. These distinct solutions relate as follows: $\nu_2 = -\nu_1$ and $\nu_4 = -\nu_3$. The four independent Floquet solutions in the most general case are thus of the form $e^{\mu_1 \t} \Phi_1(\t)$, $e^{-\mu_1 \t} \Phi_2(\t)$, $e^{\mu_3 \t} \Phi_3(\t)$ and $e^{-\mu_3 \t} \Phi_4(\t)$, where $\Phi_1, \Phi_2, \Phi_3$ and $\Phi_4$ are $2 \pi$-periodic functions. Recalling the earlier discussion in this section, we note that it is possible that there are less than four independent Floquet solutions if at least two eigenvalues $\lambda_j = e^{2 \pi i \nu_j}$ of $X(2 \pi)$ are equal. From the properties of $\Delta$, we can see that this happens in two situations. One possibility is if we have $\text{Re}[\nu_j] = n/2$ for some $n \in \mathbb{Z}$ and $\text{Im}[\nu_j] = 0$, for one of the solutions $\nu_j$. Indeed, the eigenvalues $e^{2 \pi i \nu_j}$ and $e^{-2 \pi i \nu_j}$ are equal in that case. Another possibility is if, for two solutions $\nu_j$ and $\nu_l$, we have $\text{Re}[\nu_j] = \text{Re}[\nu_l] + n$ for some $n \in \mathbb{Z}$ and $\text{Im}[\nu_j] = \text{Im}[\nu_l]$. In this case $e^{2 \pi i \nu_j}$ and $e^{2 \pi i \nu_l}$ will also be equal.

We find the following formula for $\Delta$, derived in Appendix \ref{ExtPow}:

\begin{equation} \label{DeltaFormula}
\Delta \; = \; 1 + \d\sum_{n=1}^{\infty} \l[ \d\sum_{j_1 = - \infty}^{\infty} \d\sum_{j_2 = j_1 + 2}^{\infty} \ldots \d\sum_{j_n = j_{n-1} + 2}^{\infty} \l( \d\prod_{l=1}^n  \varepsilon_{j_l} \varepsilon_{j_l +1} \r) \r] \text{.}
\end{equation}
It is clear from Eq. \eqref{epsj} that $\varepsilon_j \approx \frac{k_z V_0}{2 j^2 \o_0}$ as $\abs{j} \to \infty$. We can then try to use the fact that $\varepsilon_j$ drops off as $1/j^2$ to approximate the cumbersome formula \eqref{DeltaFormula}. If we can assume that $\abs{\text{Im}[\nu]} \ll \abs{\text{Re}[\nu]}$, then since $-0.5 < \text{Re}\l[ \nu \r] \leq 0.5$ the quantity $\nu$ becomes negligible with respect to $j$ already for quite small $\abs{j}$ in that case. In Section \ref{StabInterf}, we see that this is a good assumption in both coronal and photospheric conditions, if the Alfv\'en Mach number $M_A = (V_{0i}-V_{0e})/v_{Ai}$ is small enough.

One could then try considering, as a first approximation for $\Delta$, only the first few terms on the right-hand side of Eq. \eqref{DeltaFormula}. Taking $n=1$ and $j_1 \in \{-1,0,1\}$, this would yield:

\begin{equation} \label{DeltaApprox}
\Delta \;\; \approx \;\; 1 \;\; + \;\; \varepsilon_{-1} \; \varepsilon_0 \;\; + \;\; \varepsilon_0 \; \varepsilon_1 \text{.}
\end{equation}
Equation \eqref{DeltaApprox} corresponds to Eq. \eqref{Delta} with the infinite determinant in the right-hand side truncated to a $3 \times 3$ determinant centered on the row with $\varepsilon_0$. If one starts from the incompressible versions of Eqs. \eqref{eq1}-\eqref{eq4}, one can derive that $m=ik$ in an incompressible plasma. The approximation of truncating the series in Eq. \eqref{DeltaFormula} is only valid for large enough $\abs{j}$ and away from the poles of the $\varepsilon_j$. Under the assumption that $\abs{K} v_A \ll \o_0$ and $\abs{K} v_s \ll \o_0$, this condition is fulfilled. We note that in the solar atmosphere, $v_A$ and $v_s$ are rougly of the same order of magnitude. Since $K = k^2 - \abs{m}^2$ for surface waves, we have $K \to 0$ in the incompressible limit. Eq. \eqref{DeltaApprox}  thus gives us an approximation for $\Delta$ at least in a weakly compressible plasma, but could maybe even be correct more generally. 

Analytical solutions for $m_i$ and $m_e$ in function of $\nu$ can then be derived from Eq. \eqref{DeltaApprox}. The obtained expressions for $m_i(\nu)$ and $m_e(\nu)$ are very complicated but, introducing numerical values relevant for the physical conditions of the solar structure being considered, we can use them together with a third relation involving $\nu$, $m_i$ and $m_e$. This other relation can theoretically be the one derived in the next subsection, although in practice one of those we derive in Section \ref{StabInterf} under approximating circumstances will probably be preferred. We note that $\nu$ is the same on both sides of the interface, as will be explained in the next section. In contrast to this, $m$ is in general not identical on both sides. It can also be seen that $\Delta(\nu, m) = \Delta(\nu, -m)$. Hence, if $m$ is a solution then so is $-m$. This is reflected in the form of the spatial function $F$ in Eq. \eqref{SolF}.

\subsection{Other perturbed physical quantities}\label{timeEvolution} \label{subse}

With $h(t) = G(t) g(t)$, the compression term can be written as $\c = F(x) h(t) \exp \{ i (k_y y + k_z z) \}$. From this and Eqs. \eqref{eq1}-\eqref{eq4}, the following expressions for the perturbed quantities can then easily be derived:

\begin{align}
\xi_x &= - \d\frac{i m}{k^2 + m^2} \; \tilde{F}(x)\; h_1(t) \; \e \text{,} \label{xix}\\
\xi_y &= - \d\frac{i k_y}{k^2 + m^2} \; F(x)\; h_2(t) \; \e \text{,} \label{xiy}\\
\xi_z &= - \d\frac{i k_z}{k^2 + m^2} \; F(x)\; h_3(t) \; \e \text{,} \label{xiz}\\
B_{1x} &= B_{0z} \; \d\frac{k_z m}{k^2 + m^2} \; \tilde{F}(x) \; h_1(t) \; \e \text{,} \\
B_{1y} &= B_{0z} \; \d\frac{k_z k_y}{k^2 + m^2} \; F(x) \; h_2(t) \; \e \text{,} \\
B_{1z} &= B_{0z} \; F(x) \l( \d\frac{k_z^2}{k^2 + m^2} \; h_3(t) \; - \; h(t) \r) \e \text{,}\\
p_1 &= -\rho_0 \; v_s^2 \; F(x) \; h(t) \; \e \text{,} \\
\rho_1 &= -\rho_0 \; F(x) \; h(t) \; \e \text{,} \\
P_1 &= \rho_0 \; \l( \d\frac{k_z^2 v_A^2}{k^2 + m^2} h_3(t) \; - \; \l( v_A^2 + v_s^2 \r) h(t) \r) \notag\\ 
& \hspace{4cm} F(x) \; \e \text{,} \label{P1}
\end{align}
with $\tilde{F}(x) = C_1 e^{imx} - C_2 e^{-imx}$ and where $h_1$, $h_2$, $h_3$ are still unknown but have to satisfy 

\begin{equation}\label{h1h2h3h}
\d\frac{m^2}{k^2 + m^2} \; h_1(t) \; + \; \d\frac{k_y^2}{k^2 + m^2} \; h_2(t) \; + \; \d\frac{k_z^2}{k^2 + m^2} \; h_3(t) \; = \; h(t) \text{,}
\end{equation}
by the definition $\c = \partial \xi_x / \partial x + \partial \xi_y / \partial y + \partial \xi_z / \partial z$. Each of the expressions \eqref{xix}-\eqref{P1} is different for each region on both sides of the interface. The functions $\tilde{F}$, $F$, $h$, $h_1$, $h_2$ and $h_3$ in particular will have a different version in both regions. For the spatial functions $\tilde{F}$ and $F$, only one of the two coefficients $C_1$ and $C_2$ must be retained in each region. We make the choice that $C_1 = 0$ in the $x<0$ region, and $C_2 = 0$ in the $x>0$ region.

Writing $h_1(t) = G_1(t) g(t)$, $h_2(t) = G_2(t) g(t)$ and $h_3(t) = G_3(t) g(t)$, the following expressions for $G_1$, $G_2$ and $G_3$ can be derived from Eqs. \eqref{eq1}-\eqref{eq3}, Eqs. \eqref{xix}-\eqref{xiz} and Eq. \eqref{P1} (see Appendix \ref{AppendixG1G2G3}):

%\begin{align}
%G_1(\t) \; &= \; G_2(\t) \notag \\
%&= \; \d\frac{k^2 + m^2}{2 \o_0^2} \; e^{\mu \t} \d\sum_{j = -\infty}^{\infty} \d\frac{2 \o_0^2 \l( v_A^2 + v_s^2 \r) \varphi_j + k_z v_A^2 \psi_j}{\l(j + \nu \r)^2 \o_0^2 - k_z^2 v_A^2} \; e^{i j \t} \text{,} \\
%G_3(\t) \; &= \; -\d\frac{k^2 + m^2}{2 k_z \o_0^2} \; e^{\mu \t} \d\sum_{j = -\infty}^{\infty} \psi_j \; e^{i j \t} \text{,}
%\end{align}

\begin{align}
G_1(\t) \; &= \; G_2(\t) \notag \\
&= \; K^2 v_s^2 \; e^{\mu \t} \d\sum_{j = -\infty}^{\infty} \d\frac{1}{\l(j + \nu \r)^2 \o_0^2 - K^2 v_A^2} \; \varphi_j \; e^{i j \t} \text{,} \\
G_3(\t) \; &= \; \d\frac{K^2}{k_z^2} \; e^{\mu \t} \d\sum_{j = -\infty}^{\infty} \l(1 - \frac{\l( k_y^2+ m^2 \r) v_s^2}{\l(j + \nu \r)^2 \o_0^2 - K^2 v_A^2} \r) \; \varphi_j \; e^{i j \t} \text{.}
\end{align}
It can also be checked that Eq. \eqref{h1h2h3h} is indeed fulfilled with these expressions. Now, the following two boundary conditions have to be fulfilled at the interface between the two regions (i.e., at $x=0$) for physical reasons:

\begin{align}
[ \xi_x ] = 0 \text{,} \label{BCxi}\\
[ P_1 ] = 0 \text{,} \label{BCP1}
\end{align}
where $[f] = \lim_{x \downarrow 0} f(x) - \lim_{x \uparrow 0} f(x)$ denotes the jump in a quantity $f$ across the interface. We note that, similarly as for the frequency of a normal mode in the case without background flow, $\nu$ is the same inside and outside. It has proven to be too difficult to show mathematically, but it can be explained as follows. When a perturbation is unstable, its growth rate is determined solely by $\nu$: the growth of every perturbed quantity is namely expressed by the factor $e^{-\text{Im}[\nu]t}$. Therefore, since quantities such as $\xi_x$ and $P_1$ have to be continuous at the interface $x=0$, $\nu$ must be the same on both sides of the interface.

From the two equations \eqref{BCxi} and \eqref{BCP1} linking the interior and exterior solutions, a relation can be derived which has to be satisfied in order for the system determined by these equations to have nontrivial solutions:

\begin{equation} \label{disp}
\d\sum_{j=- \infty}^{\infty} \l( \rho_{0i}  m_e  \zeta_{A,j} \; + \; \rho_{0e}  m_i  \zeta_{B,j} \r) \; e^{ij \t} = 0 \text{,}
\end{equation}
where

%\begin{align*}
%\zeta_{A,j} &= \d\sum_{l=- \infty}^{\infty} \l[ \l(v_{Ai}^2 + v_{si}^2 \r) \varphi_{l,i} + \d\frac{k_z v_{Ai}^2}{2 \o_0^2} \psi_{l,i} \r]  \\
%& \qquad \hspace{1.8cm} \cdot \d\frac{2 \o_0^2 \l(v_{Ae}^2 + v_{se}^2 \r) \varphi_{j-l,e} + k_z v_{Ae}^2 \psi_{j-l,e}}{\l( j-l+\nu \r)^2 \o_0^2 - k_z^2 v_{Ae}^2}, \\
%\zeta_{B,j} &= \d\sum_{l=- \infty}^{\infty} \l[ \l(v_{Ae}^2 + v_{se}^2 \r) \varphi_{l,e} + \d\frac{k_z v_{Ae}^2}{2 \o_0^2} \psi_{l,e} \r]  \\
%& \qquad \hspace{1.8cm} \cdot \d\frac{2 \o_0^2 \l(v_{Ai}^2 + v_{si}^2 \r) \varphi_{j-l,i} + k_z v_{Ai}^2 \psi_{j-l,i} }{\l( j-l+\nu \r)^2 \o_0^2 - k_z^2 v_{Ai}^2} \text{.}
%\end{align*}

\begin{align*}
\zeta_{A,j} &= \d\sum_{l=- \infty}^{\infty} \frac{2 \o_0^2 v_{si}^2 v_{se}^2 \l[ \l( l+\nu \r)^2 \o_0^2 - k_z^2 v_{Ai}^2 \r] \varphi_{l,i} \varphi_{j-l,e}}{\l[ \l( l+\nu \r)^2 \o_0^2 - K_i^2 v_{Ai}^2 \r] \l[ \l(j- l+\nu \r)^2 \o_0^2 - K_e^2 v_{Ae}^2 \r]  } \text{,} \\
\zeta_{B,j} &= \d\sum_{l=- \infty}^{\infty} \frac{2 \o_0^2 v_{si}^2 v_{se}^2 \l[ \l( l+\nu \r)^2 \o_0^2 - k_z^2 v_{Ae}^2 \r] \varphi_{l,e} \varphi_{j-l,i}}{\l[ \l( l+\nu \r)^2 \o_0^2 - K_e^2 v_{Ae}^2 \r] \l[ \l(j- l+\nu \r)^2 \o_0^2 - K_i^2 v_{Ai}^2 \r]  }   \text{.}
\end{align*}
Equation \eqref{disp}, arising from the boundary conditions at the interface, is usually called the dispersion relation when there is no oscillating background flow. Both $m_i$ and $m_e$ being determined by Eq. \eqref{Delta} (by inserting respectively interior and exterior values for the background quantities), Eq. \eqref{disp} determines the only remaining unknown, $\nu$.

From Eq. \eqref{disp}, the following has to hold for every $j \in \mathbb{Z}$:

\begin{equation} \label{mRatios}
\d\frac{m_e}{m_i} = -\d\frac{\rho_{0e} \zeta_{B,j}}{\rho_{0i} \zeta_{A,j}} \text{.}
\end{equation}
For $j = 0$ this gives us the relation

\begin{equation} \label{mRatiosj0}
\d\frac{m_e}{m_i} = - \d\frac{\rho_{0e}}{\rho_{0i} } \d\frac{\d\sum_{l=- \infty}^{\infty} \frac{ \l[ \l( l+\nu \r)^2 \o_0^2 - k_z^2 v_{Ae}^2 \r] \varphi_{l,e} \varphi_{-l,i}}{\l[ \l( l+\nu \r)^2 \o_0^2 - K_e^2 v_{Ae}^2 \r] \l[ \l(l-\nu \r)^2 \o_0^2 - K_i^2 v_{Ai}^2 \r] } }{ \d\sum_{l=- \infty}^{\infty} \frac{\l[ \l(l -\nu \r)^2 \o_0^2 - k_z^2 v_{Ai}^2 \r] \varphi_{l,e} \varphi_{-l,i}}{\l[ \l( l+\nu \r)^2 \o_0^2 - K_e^2 v_{Ae}^2 \r] \l[ \l(l-\nu \r)^2 \o_0^2 - K_i^2 v_{Ai}^2 \r] } } \text{.}
\end{equation}
We note that this is not an explicit solution for $m_e/m_i$, because both $m_i$ and $m_e$ appear on the right-hand side, through $K_i$ and $K_e$ respectively. While Eq. \eqref{mRatiosj0} seems difficult to use directly, it gives us some information. Indeed, we learn from this equation that, whereas in the incompressible case we have $m_e/m_i = 1$, in the compressible case $m_e/m_i$ seems to depend on $k_z$ as well as $k_y$, the latter appearing only in $K_i$ and $K_e$. In the next section, we see that the only perturbation quantities the stability of the interface depends on are $k_z$ and $m_e/m_i$. This means that, while the abstraction is made that only the longitudinal wavenumber $k_z$ is important for the stability of the interface in an incompressible plasma, the perpendicular wavenumber $k_y$ seems to have some influence as well in the more realistic case of a compressible plasma.

%One way the previous derivations can be put to use under fixed physical conditions is by determining $m_i$ and $m_e$ as a function of $\nu$ from the approximation Eq. \eqref{DeltaApprox}, and insert this in an approximate version of Eq. \eqref{mRatiosj0}. Indeed, the terms in both sums on the right-hand side of Eq. \eqref{mRatiosj0} are $o(1/\abs{l}^4)$ as $\abs{l} \to \infty$ (Fourier theory namely ensures the coefficients $\varphi_l$ are $o(1/\abs{l})$ as $\abs{l} \to \infty$ if $G$ is differentiable). One can then try to approximate the sums in Eq. \eqref{mRatiosj0} by truncating the series to the terms with $l=0$, $1$ and $-1$ for example. This would be acceptable under the assumption that, for both regions, $\abs{K} v_s \ll \o_0$, $\abs{K} v_s \ll \o_0$ (which are the same assumptions under which we found Eq. \eqref{DeltaApprox} to be a good approximation) and $\abs{\l( l \pm \nu \r)^2 \o_0^2 - k_z^2 v_{A}^2} \lessapprox 1$. In the next section it will become clear that for coronal loop conditions this last condition is fulfilled for $\abs{l} > 1$, thus making this approximation also acceptable in that case. The Fourier coefficients $\varphi_{-1}$, $\varphi_0$ and $\varphi_1$ of $G$ need to be computed numerically.

\section{Stability of the interface} \label{StabInterf}

\subsection{Governing equation}

The results derived in the preceding sections permit us to say something about the stability of the interface at the structure's boundary. This stability is determined by the temporal evolution of the normal displacement $\xi_x$, which is governed by the following equation:

\begin{equation} \label{Eqxix}
\d\frac{D^2 \xi_x}{Dt^2}  \; + \; k_z^2 v_A^2 \xi_x \; = \; \frac{-1}{\rho_0} \frac{\pa P_1}{\pa x} \text{.}
\end{equation}
There is again a different version of Eq. \eqref{Eqxix} for each region, namely for $x<0$ and for $x>0$. For each of the two regions, the respective versions of expression \eqref{P1} for $P_1$ can then be inserted in the respective versions of Eq. \eqref{Eqxix}. These can then be put together by expressing $C_1$ in function of $C_2$ thanks to Eq. \eqref{BCP1}, and yield the following equation governing the displacement of the interface over time:

\begin{align}
&\l( \rho_{0e} m_i \d\frac{D_e^2 \l(\xi_x\bigr\rvert_{x=0} \r)}{Dt^2} \; + \; \rho_{0i} m_e \frac{D_i^2 \l(\xi_x\bigr\rvert_{x=0} \r)}{Dt^2} \r) \notag\\
& \hspace{1.5cm} + \;  k_z^2 \l( \rho_{0e} v_{Ae}^2 m_i \xi_x\bigr\rvert_{x=0} \; + \; \rho_{0i} v_{Ai}^2 m_e \xi_x\bigr\rvert_{x=0} \r) = 0 \text{,} \label{xixie}
\end{align}
where $D_i/Dt = \pa / \pa t + i k_z V_{0i} \sin (\o_0 t)$, $D_e/Dt = \pa / \pa t + i k_z V_{0e} \sin (\o_0 t)$, and $\xi_x\bigr\rvert_{x=0}$ is $\xi_x$ evaluated at $x=0$.

Using a similar trick as was used for $\c$ before and which was also used by \citet{BarbulescuEtAl2019} in a different setup, we write $\xi_x\bigr\rvert_{x=0}$ as 

\begin{equation}\label{trick}
\xi_x\bigr\rvert_{x=0}(t) = \eta(t) \exp \l\{\frac{- i A \sin \l( \o_0 t \r)}{\o_0} \r\} \text{,}
\end{equation}
with $A = k_z \frac{V_{0e} m_i \rho_{0e} + V_{0i} m_e \rho_{0i}}{m_i \rho_{0e} + m_e \rho_{0i}}$. This assumption changes nothing to the stability of $\xi_x\bigr\rvert_{x=0}(t)$ since the moduli of $\xi_x\bigr\rvert_{x=0}(t)$ and $\eta(t)$ are the same for every $t \in \mathbb{R}^+$, and therefore $\xi_x\bigr\rvert_{x=0}$ is unbounded if and only if $\eta$ is unbounded.   Inserting Eq. \eqref{trick} in Eq. \eqref{xixie} greatly simplifies the equation, which can be worked out to yield the following ODE for $\eta$:

\begin{equation}\label{Mathieu}
\d\frac{d^2 \eta(\tau)}{d \tau^2} + \l( a - 2 q \cos \l( 2 \tau \r) \r) \eta(\tau) = 0 \text{,}
\end{equation}
with $\tau = \o_0 t$ and

%\begin{align}
%a &= \d \frac{k_z^2}{ \o_0^2 \l( \rho_{0i} m_e + \rho_{0e} m_i \r)^2} \l\{ m_i^2 \rho_{0e}^2 v_{Ae}^2 +  m_e^2 \rho_{0i}^2 v_{Ai}^2 \r. \notag \\
%& \hspace{1.5cm} \l. - m_i m_e \rho_{0i} \rho_{0e} \l[ \frac{1}{2} \l(V_{0i} - V_{0e} \r)^2 -  \l( v_{Ai}^2 + v_{Ae}^2 \r) \r] \r\} \text{,} \label{a1} \\
%q &= \d\frac{k_z^2 \rho_{0i} \rho_{0e} m_i m_e \l( V_{0e} - V_{0i} \r)^2}{4 \o_0^2 \l( \rho_{0i} m_e + \rho_{0e} m_i \r)^2} \text{.} \label{q1}
%\end{align}

\begin{align}
a &= \d\frac{k_z^2}{\o_0^2} \d\frac{\rho_{0i} v_{Ai}^2 m_e + \rho_{0e} v_{Ae}^2 m_i}{\rho_{0i} m_e + \rho_{0e} m_i} - \frac{k_z^2 \rho_{0i} \rho_{0e} m_i m_e \l(V_{0i} - V_{0e} \r)^2}{2 \o_0^2 \l( \rho_{0i} m_e + \rho_{0e} m_i \r)^2} \text{,} \label{a1} \\
q &= \d\frac{k_z^2 \rho_{0i} \rho_{0e} m_i m_e \l( V_{0i} - V_{0e} \r)^2}{4 \o_0^2 \l( \rho_{0i} m_e + \rho_{0e} m_i \r)^2} \text{.} \label{q1}
\end{align}
The parameters $a$ and $q$ can be rewritten in normalized form as follows: 

%\begin{align}
%a &= \d\frac{\kappa_z^2 \l( r^2 \overline{v}_A^2 + \overline{m}^2 + \overline{m} r \l( 1 + \overline{v}_A^2 - \frac{1}{2} M_A^2 \r) \r) }{\l( \overline{m} + r \r)^2} \text{,} \label{a2}\\
%q &= \d\frac{\kappa_z^2 r \overline{m} M_A^2}{4 \l( \overline{m} + r \r)^2} \label{q2} \text{,}
%\end{align}

\begin{align}
a &= \kappa_z^2 \l[ \d\frac{\ov{m} + r \ov{v}_A^2}{\ov{m} + r} - \frac{r \ov{m} M_A^2}{2 \l( \ov{m} + r \r)^2} \r] \text{,} \label{a2}\\
q &= \kappa_z^2 \d\frac{ r \overline{m} M_A^2}{4 \l( \overline{m} + r \r)^2} \label{q2} \text{,}
\end{align}
where we introduced the normalized quantities $\kappa_z = k_z v_{Ai} / \o_0$, $\overline{v}_A = v_{Ae} / v_{Ai}$, $r = \rho_{0e} / \rho_{0i}$, $\overline{m} = m_e/m_i$ and $M_A = (V_{0i}-V_{0e}) / v_{Ai}$. 

We notice that the first term of $a$ consists of an expression that resembles the squared kink frequency and which we call the pseudo squared kink expression. The second term of $a$ resembles a Doppler-shifting correction and equals $-2q$. In an incompressible plasma, the first term would be the squared kink frequency (since then $m_i, m_e \to k$). We also note that if there is no background flow (i.e., $V_{0i} = V_{0e} = 0$), we are left with only the pseudo squared kink expression in the factor in front of $\eta(\t)$ in Eq. \eqref{Mathieu}. The solution is then a surface wave with frequency $\nu$ determined by the dispersion relation

\begin{equation} \label{nuConstFlow}
\nu^2 = \kappa_z^2 \d\frac{\ov{m} + r \ov{v}_A^2}{\ov{m} + r} \text{,}
\end{equation}
where $m_i$ and $m_e$ are determined in function of $\nu$ through their respective version of Eq. \eqref{DeltaFormula}. Eq. \eqref{nuConstFlow} is a transcendental equation in $\nu$ with more than one solution, which means that there are several different modes in a compressible plasma. This is in contrast with an incompressible plasma, where there is only one mode, for which the expression of the frequency is readily readable from Eq. \eqref{nuConstFlow} and is equal to the kink frequency. These results about modes at an interface in a compressible plasma without background flow are already known (see for example \citet{Priest2014}).

\subsection{Instability conditions}

\begin{figure}
   \centering
   \includegraphics[scale=0.165]{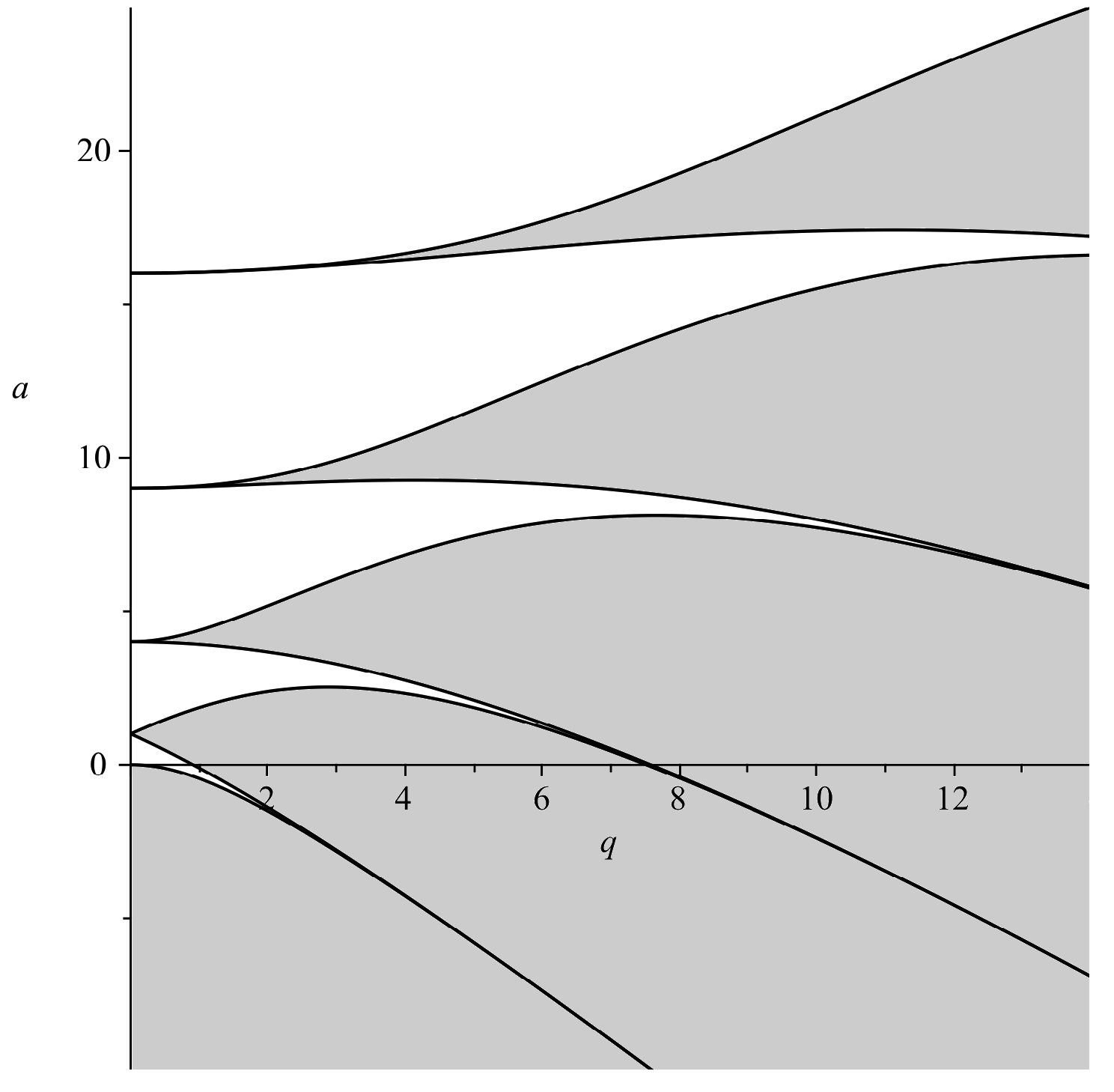}
      \caption{Stability diagram of the Mathieu equation. The white zones represent regions of the parameter space where the solution is stable, whereas the gray zones represent regions of the parameter space where the solution is unstable.}
         \label{StabDiag}
   \end{figure}

Equation \eqref{Mathieu} is the Mathieu equation, which also pertains to the class of equations obeying Floquet theory and which has been extensively studied for example by \citet{McLachlan1947}. It does not have an analytical solution but the stability of its solution in function of its parameters is well-known. The stability of the solutions of the Mathieu equation \eqref{Mathieu} in function of the parameters $a$ and $q$ is often represented in the so-called stability diagram (see Figure \ref{StabDiag}). The white zones in the figure are regions where the solutions are stable, whereas the gray zones are region where the solutions are unstable.

\subsubsection{Constant shear flow}

In order to gain some insight on the nature of the potential instabilities, we follow \citet{HillierEtAl2019}, who studied the stability of an interface in a similar model. We start by looking at the case with a constant background shear flow (i.e., we assume $\o_0 = 0$). We have to modify the procedure above a little bit in order to be mathematically correct: instead of Eq. \eqref{trick}, we assume $\xi_x\bigr\rvert_{x=0}(t) = \eta(t) \exp \l\{- i A t \r\}$ in this particular case. The obtained equation is then

\begin{equation}\label{PseudoMathieu}
\d\frac{d^2 \eta(\tau)}{d \tau^2} + \kappa_z^2 \l[ \d\frac{\ov{m} + r \ov{v}_A^2}{\ov{m} + r} - \frac{r \ov{m} M_A^2}{\l( \ov{m} + r \r)^2} \r]\; \eta(\tau) = 0 \text{.}
\end{equation}
It has normal mode solutions with normalized frequency $\nu$ defined by the dispersion relation

\begin{equation} \label{DispMat1}
\nu^2 = \kappa_z^2 \l[ \d\frac{\ov{m} + r \ov{v}_A^2}{\ov{m} + r} - \frac{r \ov{m} M_A^2}{\l( \ov{m} + r \r)^2} \r] \text{.}
\end{equation}
This transcendental equation in $\nu$ has multiple solutions as well, and thus we see that there are also different modes in the case of a constant shear flow in a compressible plasma. A particular mode will be unstable to the KHI if 

\begin{equation} \label{instCondConst}
M_A^2 > \d\frac{\l( \ov{m} + r \r) \l( \ov{m} + r \ov{v}_A^2  \r)}{r \ov{m}}
\end{equation}
is satisfied. Different modes will thus have different instability conditions in a compressible plasma. If we take the incompressible limit, the above derivations match the already well-known results of a constant shear flow in an incompressible plasma (see for example \citet{Chandrasekhar1961}).

\subsubsection{Oscillating shear flow}

In the case of an oscillatory background flow ($\o_0 \neq 0$), Eq. \eqref{Mathieu} governs the evolution of the normal displacement at the interface over time. In the same way as for Eq. \eqref{EqG}, Floquet theory allows us to write an independent solution to the Mathieu equation as follows \citep{McLachlan1947}:

\begin{equation}\label{seriesMathieu}
\eta(\t) \; = \; e^{i \nu \t} \d\sum_{j = -\infty}^{\infty} \phi_j \; e^{2 i j \t} \text{.}
\end{equation}
This resembles a normal mode with frequency $\nu$, but with a periodic function replacing the constant. One can then find the following recursion relation by inserting Eq. \eqref{seriesMathieu} into Eq. \eqref{Mathieu}:

\begin{equation} \label{recursionMathieu}
\epsilon_j \; \phi_{j-1} \;\; + \;\; \phi_j \;\; + \;\; \epsilon_j \; \phi_{j+1} \;\; = \;\; 0\text{,} 
\end{equation}
where

\begin{equation}
\epsilon_j = \d\frac{q}{\l( 2j + \nu \r)^2-a} \text{.}
\end{equation}
If the denominator of one of the $\epsilon_j$ vanishes, there can be a resonance between the background oscillator and the induced surface waves.

Now some analytical progress can be made in the limiting case $q \to 0$, corresponding to a small squared shear rate $k_z^2 (V_{0i}-V_{0e})^2$ with respect to the squared frequency $\o_0^2$. Indeed, for $q \ll 1$, we see from Eq. \eqref{recursionMathieu} that in order to have a nontrivial solution \eqref{seriesMathieu} we need the denominator of at least one of the $\epsilon_j$ to vanish. For that, we need to have

\begin{equation} \label{res}
\l( 2j + \nu \r)^2-a \approx 0
\end{equation}
for a $j \in \mathbb{Z}$. In case $\nu^2 \approx a$, that is to say, if

\begin{equation} \label{j0}
\nu^2 \approx \kappa_z^2 \l[ \d\frac{\ov{m} + r \ov{v}_A^2}{\ov{m} + r} - \frac{r \ov{m} M_A^2}{2 \l( \ov{m} + r \r)^2} \r] \text{,}
\end{equation}
then Eq. \eqref{res} is fulfilled for $j = 0$. Since Eq. \eqref{j0} is again a transcendental equation in $\nu$ with multiple solutions, there are multiple modes possible. These modes are surface waves with frequency $\nu$. Such a mode will be unstable if 

\begin{equation} \label{instCondOsc}
M_A^2 > 2 \d\frac{\l( \ov{m} + r \r) \l( \ov{m} + r \ov{v}_A^2  \r)}{r \ov{m}}
\end{equation} 
is satisfied. This renders the right-hand side in Eq. \eqref{j0}, which is $a$, negative. Hence, for $q \ll 1$, the region of the parameters space which corresponds to the KHI is $a<0$. It can be seen on Figure \ref{StabDiag} that this region is indeed uniformly unstable. We note that the right-hand side in Eq. \eqref{instCondOsc} is a factor of $2$ larger than in Eq. \eqref{instCondConst}. In a first approximation, we could thus evaluate the minimum value of $M_A$ for the KHI to develop in the case of an oscillatory shear flow with $q \ll 1$ to be $\sqrt{2}$ times higher than in the case of a constant shear flow with $q \ll 1$. This is assuming that the corresponding values of $m_i$ and $m_e$ are equal in both cases. In an incompressible plasma, this is obviously correct. In a compressible plasma, however, there is no reason to suggest a priori that this is the case in general. Nevertheless, the values might still be close, such as in a weakly compressible plasma for example.

If on top of Eq. \eqref{res} being satistfied for $j=0$ it is also satisfied for another $j \in \mathbb{Z}$, then there is a resonance between a mode satisfying Eq. \eqref{j0} and the background oscillator. From $\nu^2 = a$ and  Eq. \eqref{res} with $j\neq0$, this means that we must have

\begin{equation}
a = j^2 \text{.}
\end{equation}
In Figure \ref{StabDiag} we see the regions of resonance instability around $a=j^2$, with $j \in \mathbb{Z}_0$. Hence, for $q \ll 1$, the KHI and the resonance instability are clearly distinct features and are mutually exclusive.

As was also mentionned by \citet{HillierEtAl2019} (themselves based on \citet{Bender&Orszag1999}), the dominant resonance in this limit is the one with $j = 1$. Its instability reagion is borded by the curves

\begin{equation}
a = j^2 \pm q + O(q^2)  \hspace{1cm} \text{ as } q \to 0
\end{equation}
and its maximum growth rate is

\begin{equation}
\text{Im}[\nu] = \d\frac{q}{2} + O(q^2)  \hspace{1cm} \text{ as } q \to 0 \text{.}
\end{equation}
We could thus express this growth rate through the equation $\text{Im}[\nu] = \frac{q}{2}$ as an approximation. This is an implicit equation in $\nu$, because $q$ depends on it through $\ov{m}$.

\subsection{Solar applications: Relevant region of the parameter space}

From Eqs. \eqref{a2} and \eqref{q2}, we see the parameters $a$ and $q$ are related as follows:

\begin{equation} \label{a=f(q)}
a = \l\{ \d\frac{4 \l( \ov{m} + r \r) \l( \ov{m} + r \ov{v}_A^2  \r)}{r \ov{m} M_A^2} - 2 \r\} \; q \text{.}
\end{equation}
In the incompressible limit, $\ov{m} \to 1$ and we find the same relation as \citet{BarbulescuEtAl2019} for their model. For fixed values of the background quantities and for a fixed value of $\overline{m}$, Eq. \eqref{a=f(q)} represents a line in the $aq$-plane. We note, however, that changing the value of $k_z$ and/or $k_y$ a priori changes the value of $\ov{m}$ as well and that hence the modes do not lie all on the same line. This is unlike the case of an incompressible plasma, where one stays on the same line when changing $k_z$, as was already found for example by \citet{BarbulescuEtAl2019} in a similar model. 

We find that the slope in Eq. \eqref{a=f(q)} is minimized for $\ov{m} = r \ov{v}_A$, in which case the line's equation is

\begin{equation} \label{minSlope}
a= \l\{ \frac{ 4 \l( 1+ \overline{v}_A \r)^2 }{M_A^2} - 2 \r\} \; q \text{.}
\end{equation}
For each set of fixed background quantities, the only physically possible values of $a$ and $q$ thus lie in the region of the stability diagram between the positive $a$-axis and the line defined by Eq. \eqref{minSlope}. We see that, in this model, it is not physically possible to have values of $(q,a)$ in the region below the line through the origin and with slope $-2$.

\subsubsection{Coronal loops}

A first application to a solar atmospheric structure would be a coronal loop, where slow waves occur near footpoints \citep{DeMoortelEtAl2002} or during flares \citep{WangEtAl2002, KumarEtAl2013}. For a standing slow wave under realistic coronal loop conditions, for which we took $(v_{Ae}, v_{si}, v_{se}) = (5/2,1/2,1/4) v_{Ai}$, the line of minimum slope, Eq. \eqref{minSlope}, lies almost on the positive $a$-axis. Indeed, with $\overline{v}_A = 5/2$ and for $M_A = 0.066$ for example, we find the line of minimum slope to be given by $a = 11415q$. If we assume incompressibility and fix $\ov{m}=1$, Eq. \eqref{a=f(q)} defines a line as well and we find $a=12486q$. For the less realistic value of $M_A = 1$, we find that the line of minimum slope is given by $a=47q$ whereas in the incompressible limit the line \eqref{a=f(q)} becomes $a=52q$. As we can see from Figure \ref{StabDiag}, the region of the $aq$-diagram relevant for coronal loop conditions is thus stable with respect to the KHI. It could be unstable with respect to resonance; however, since in reality a finite cylindrical structure is closed both azimuthally and longitudinally, the respective wavenumbers will be entire and thus take on discrete values. In the $aq$-diagram, the possible values will thus be a set of discrete points $(a_n, q_n)$. Seeing how the vast majority of the area is a stable region in the relevant region between the positive $a$-axis and the minimum-slope line \eqref{minSlope}, it is likely that this parametric resonance instability is avoided as well.

\begin{table}[!htb]
\begin{tabular}{|l|l|l|l|}
\hline
\textbf{conditions} & \textbf{$M_A$} & \textbf{minimum slope} & \textbf{incomp. slope} \\ \hline
coronal             & 0.066          & 11415                  & 12486                  \\ \hline
 photospheric        & 0.066          & 1450                   & 1615                   \\ \hline
 photospheric        & 0.36          & 46                      & 52                    \\ \hline
 coronal             & 1              & 47                     & 52                     \\ \hline
photospheric        & 1              & 4.25                   & 5                      \\ \hline
\end{tabular}
\caption{Table summarizing the minimum slope in Eq. \eqref{minSlope} and the slope of Eq. \eqref{a=f(q)} in the incompressible limit, for different atmospheric conditions. The first two rows are for realistic values of the longitudinal velocity for a slow surface wave on a discontinuous boundary of a cylindrical structure. The third row is for realistic values of the longitudinal velocity around the resonant position for a resonant slow wave in a photospheric pore. The last two lines are examples where $M_A$ begins to be unrealistically high and which show that even then the slopes are not low enough to be in an unstable regime. For coronal conditions we took $(v_{Ae}, v_{si}, v_{se}) = (5/2,1/2,1/4) v_{Ai}$, whereas for photospheric conditions we took $(v_{Ae}, v_{si}, v_{se}) = (1/4,1/2,3/4) v_{Ai}$.}
\label{tab:my-table}
\end{table}

We conclude that the oscillating shear velocity of a standing slow wave in a coronal loop would not trigger the KHI on its own, without the involvement of other physical processes. This is in clear contrast with the fast kink waves. Indeed, according to the incompressible models of \citet{BarbulescuEtAl2019} and \citet{HillierEtAl2019}, which are similar to this model but where the velocity and magnetic fields are perpendicular instead of parallel, fast kink waves are predicted to excite the KHI on the loop boundary. This has also been confirmed by numerical simulations.

\subsubsection{Photospheric pores}

As was mentionend in the introduction, propagating slow waves have also been observed in photospheric pores. One would be curious whether standing versions of these modes could trigger the KHI under these conditions. The model can be used to examine the stability of slow waves in these structures as well. However, the qualitative conclusion is the same as in the case of coronal loops: with photospheric pore conditions set to $(v_{Ae}, v_{si}, v_{se}) = (1/4,1/2,3/4) v_{Ai}$, we have $\overline{v}_A = 1/4$, and, taking $M_A = 0.066$, the minimum slope is $1450$ whereas for the slope of the line defined by Eq. \eqref{a=f(q)} is  $1615$ in the incompressible limit. As a comparison, even for the unrealistic value of $M_A=1$, the minimum slope is $4.25$ and the slope in Eq. \eqref{a=f(q)} in the incompressible limit is $5$. These values are an order of magnitude lower then for their coronal counterpart, but looking at Figure \ref{StabDiag} we see that the same conclusion remains: every possible pair of parameters $(a,q)$ lie in a region that does not permit the development of the KHI. Similarly as for a coronal loop in the previous subsection, it is likely that the parametric instability arising from resonance with the driver is also avoided and that the pore therefore remains stable.

Additionally, one could investigate the stability of resonant slow waves, which can occur in the presence of a transition layer at the pore's boundary. Indeed, slow surface waves are known to be resonantly absorbed in photospheric conditions. It is worth wondering whether the great longitudinal velocity shear that arises around the resonant point could be enough to trigger the KHI. In ideal MHD, the longitudinal velocity of resonant slow modes displays a hyperbolic singularity \citep{SakuraiEtAl1991}, but in the presence of finite resistivity this singularity disappears to make place for a sharp but continuous profile. \citet{Kovitya&Cram1983} reported values of electrical resistivity in sunspots, with a minimum resistivity corresponding to a magnetic Reynolds number of $10^7$. \citet{Erdelyi1997} derived an expression for the longitudinal Lagrangian displacement in the dissipative layer around the cusp resonance in the presence of finite electrical resistivity (see Eq. (30) therein). Assuming a sinusoidal transition profile with width $l=0.1R$ (where $R$ is the radius of the pore) in the squared cusp and sound speeds, a ratio of external to internal magnetic field of $B_{0ze}/B_{0zi}=0.33$, a cusp resonant position at $r_C=0.955R$ (based on the numerical computations of \citet{GoossensEtAl2021}), a value for the perturbed total pressure at the resonant position equal to its value on the interface for the corresponding surface mode in the absence of the transition layer, a real part of the frequency equal to the frequency of the corresponding surface mode in the absence of the transition layer, and a magnetic Reynolds number of $10^7$ to have a lowerbound on resistivity and thus an upperbound on realistic values for the longitudinal velocity at the resonance, we found a value of $M_A=0.36$ around the cusp resonance point for a sausage mode with $k_z R = 2$. This value of $k_z R$ is within the range of validity for the longitudinal wavenumber of the observed slow surface sausage mode in a photospheric pore by \citet{GrantEtAl2015}. The value found for $M_A$ implies a minimum slope of $46$ in Eq. \eqref{minSlope} and a slope of $52$ for Eq. \eqref{a=f(q)} in the incompressible limit. These values are still well above the values for the slopes found for $M_A=1$ in photospheric conditions. Through the model derived in this paper, we conclude from this that even resonant slow waves do not display a large enough velocity shear for an instability to develop.

\subsection{Limitations of the model}

We point out that the local model discussed in this paper is only valid under certain conditions. Firstly, the azimuthal wavenumber of the perturbation must be large, such that the azimuthal direction can be approximated by the cartesian direction of $y$. This means that we must have $k_y R \gg 1$, with $R$ the structure's radius. Secondly, for the amplitude of the oscillating background velocity to be assumed constant, the longitudinal wavelength of the perturbation must be small with respect to the structure's length $L$. We must thus have $k_z L \gg 1$. Lastly, we note that our model might not be suitable in the presence of a smooth transition layer between the interior and the exterior of the structure. Indeed, resonant absorption of the standing slow wave in the background could occur in that case \citep{YuEtAl2017b}, which leads to a steep and continuous variation in the longitudinal component of the velocity along the $x$ direction. Although one can use the model to estimate the effect of the logitudinal velocity shear in resonant absorption on the stability of the interface, as we did in last subsection, one must keep in mind that certain assumptions on which the model is based are not fulfilled in the presence of a transition layer. Firstly, there is no true discontinuous interface and, secondly, the assumption of a constant amplitude profile along $x$ for the oscillating background velocity is violated.

\section{Conclusion}

In this paper, we developed an analytical model for the local stability of a cylindrical solar atmospheric structure harboring a standing slow wave. We assumed that the structure is a straight cylinder with a circular base, of which the boundary is an interface discontinuously separating two homogeneous and compressible plasmas. The magnetic fields on both sides were assumed to be constant, straight and aligned with the interface. The velocity fields were assumed to be oscillating in time but spatially constant, in order to model the standing slow wave in the background.

We used linearized MHD to derive an equation for the compression term, $\c$. The expressions for all the other perturbed quantities could then be expressed in terms of the solution of $\c$. The spatial part of the solution along the direction normal to the interface could be analytically found to be a normal mode, with wavenumber $m$. In contrast to the incompressible approximation, the value of this wavenumber $m$ in a compressible plasma is different on both sides of the interface. We saw that $m_e/m_i$ is dependent on both $k_z$ and $k_y$, which entails that the stability of the interface depends on $k_y$ as well as $k_z$. This is in contrast with the incompressible version of this model, in which only $k_z$ has an influence on said stability.

We then found that the governing equation for the displacement component of the perturbation normal to the interface is a Mathieu equation. The stability of the solution of this equation in function of the different involved parameters being known from the litterature, we were able to describe two kinds of instabilities that could theoretically arise in the presence of an oscillating shear background flow. As an application to the solar atmosphere, we found that the physically relevant region of the parameter space corresponds to an interface that is locally stable with respect to the KHI, both in a coronal loop and in a photospheric pore. Although the interface can be unstable to a parametric resonance between the background slow wave and the induced perturbations on the interface, we concluded that it is unlikely to happen in reality. Even in the case of resonance in the cusp continuum, we concluded from our model that the longitudinal velocity shear of the resonant slow waves is not enough to trigger an instability.

We ended by noting that this model is only applicable under specific conditions. It is indeed a local model, in the sense that the spatial scale of azimuthal variations must be small compared to the structure's radius and the spatial scale of longitudinal variations must be small with respect to the loop's length. Furthermore, for a structure with a smooth transition layer the assumptions of a discontinuous transition at the boundary and a constant amplitude for the background oscillating velocity fields would be violated. This needs to be kept in mind when using this model to infer stability conditions around resonances.

\begin{acknowledgements}
This research was supported by the European Research Council (ERC) under
the European Union's Horizon 2020
research and innovation program (TVD via grant agreement No 724326),
which MG gratefully acknowledges for his Ph.D. studentship.
\end{acknowledgements}


\bibliographystyle{aa}
\bibliography{biblio}

\begin{thebibliography}{77}
\expandafter\ifx\csname natexlab\endcsname\relax\def\natexlab#1{#1}\fi

\bibitem[{{Afanasyev} {et~al.}(2019){Afanasyev}, {Karampelas}, \& {Van
  Doorsselaere}}]{AfanasyevEtAl2019}
{Afanasyev}, A., {Karampelas}, K., \& {Van Doorsselaere}, T. 2019, \apj, 876,
  100

\bibitem[{{Antolin} {et~al.}(2017){Antolin}, {De Moortel}, {Van Doorsselaere},
  \& {Yokoyama}}]{AntolinEtAl2017}
{Antolin}, P., {De Moortel}, I., {Van Doorsselaere}, T., \& {Yokoyama}, T.
  2017, \apj, 836, 219

\bibitem[{{Antolin} {et~al.}(2015){Antolin}, {Okamoto}, {De Pontieu},
  {Uitenbroek}, {Van Doorsselaere}, \& {Yokoyama}}]{AntolinEtAl2015}
{Antolin}, P., {Okamoto}, T.~J., {De Pontieu}, B., {et~al.} 2015, \apj, 809, 72

\bibitem[{{Antolin} {et~al.}(2014){Antolin}, {Yokoyama}, \& {Van
  Doorsselaere}}]{AntolinEtAl2014}
{Antolin}, P., {Yokoyama}, T., \& {Van Doorsselaere}, T. 2014, \apjl, 787, L22

\bibitem[{{Arregui}(2015)}]{Arregui2015}
{Arregui}, I. 2015, Philosophical Transactions of the Royal Society of London
  Series A, 373, 20140261

\bibitem[{{Barbulescu} {et~al.}(2019){Barbulescu}, {Ruderman}, {Van
  Doorsselaere}, \& {Erd{\'e}lyi}}]{BarbulescuEtAl2019}
{Barbulescu}, M., {Ruderman}, M.~S., {Van Doorsselaere}, T., \& {Erd{\'e}lyi},
  R. 2019, \apj, 870, 108

\bibitem[{Bender \& Orszag(1999)}]{Bender&Orszag1999}
Bender, C. \& Orszag, S. 1999, Advanced Mathematical Methods for Scientists and
  Engineers I: Asymptotic Methods and Perturbation Theory, Advanced
  Mathematical Methods for Scientists and Engineers (Springer)

\bibitem[{{Berghmans} \& {Clette}(1999)}]{Berghmans&Clette1999}
{Berghmans}, D. \& {Clette}, F. 1999, \solphys, 186, 207

\bibitem[{{Cadez} {et~al.}(1997){Cadez}, {Csik}, {Erdelyi}, \&
  {Goossens}}]{CadezEtAl1997}
{Cadez}, V.~M., {Csik}, A., {Erdelyi}, R., \& {Goossens}, M. 1997, \aap, 326,
  1241

\bibitem[{Cesari(1963)}]{Cesari1971}
Cesari, L. 1963, Asymptotic Behavior and Stability Problems in Ordinary
  Differential Equations (Springer-Verlag)

\bibitem[{{Chandrasekhar}(1961)}]{Chandrasekhar1961}
{Chandrasekhar}, S. 1961, {Hydrodynamic and hydromagnetic stability}

\bibitem[{{Chen} {et~al.}(2018){Chen}, {Li}, {Shi}, \& {Yu}}]{ChenEtAl2018}
{Chen}, S.-X., {Li}, B., {Shi}, M., \& {Yu}, H. 2018, \apj, 868, 5

\bibitem[{{Chen} {et~al.}(2020){Chen}, Li, Van~Doorsselaere, Goossens, \&
  Geeraerts}]{ChenEtAl2020}
{Chen}, S.~X., Li, B., Van~Doorsselaere, T., Goossens, M., \& Geeraerts, M.
  2020, \apj, submitted

\bibitem[{Chicone(2008)}]{Chicone2008}
Chicone, C. 2008, Ordinary Differential Equations with Applications, Texts in
  Applied Mathematics (Springer New York)

\bibitem[{Conway(1990)}]{Conway1990}
Conway, J. 1990, A Course in Functional Analysis, Graduate texts in mathematics
  (Springer)

\bibitem[{{De Moortel} \& {Hood}(2003)}]{DeMoortel&Hood2003}
{De Moortel}, I. \& {Hood}, A.~W. 2003, \aap, 408, 755

\bibitem[{{De Moortel} {et~al.}(2000){De Moortel}, {Ireland}, \&
  {Walsh}}]{DeMoortelEtAl2000}
{De Moortel}, I., {Ireland}, J., \& {Walsh}, R.~W. 2000, \aap, 355, L23

\bibitem[{{De Moortel} {et~al.}(2002){De Moortel}, {Ireland}, {Walsh}, \&
  {Hood}}]{DeMoortelEtAl2002}
{De Moortel}, I., {Ireland}, J., {Walsh}, R.~W., \& {Hood}, A.~W. 2002,
  \solphys, 209, 61

\bibitem[{{De Moortel} {et~al.}(2016){De Moortel}, {Pascoe}, {Wright}, \&
  {Hood}}]{DeMoortelEtAl2016}
{De Moortel}, I., {Pascoe}, D.~J., {Wright}, A.~N., \& {Hood}, A.~W. 2016,
  Plasma Physics and Controlled Fusion, 58, 014001

\bibitem[{{Dorotovi{\v{c}}} {et~al.}(2014){Dorotovi{\v{c}}}, {Erd{\'e}lyi},
  {Freij}, {Karlovsk{\'y}}, \& {M{\'a}rquez}}]{DorotovicEtAl2014}
{Dorotovi{\v{c}}}, I., {Erd{\'e}lyi}, R., {Freij}, N., {Karlovsk{\'y}}, V., \&
  {M{\'a}rquez}, I. 2014, \aap, 563, A12

\bibitem[{{Dorotovi{\v{c}}} {et~al.}(2008){Dorotovi{\v{c}}}, {Erd{\'e}lyi}, \&
  {Karlovsk{\'y}}}]{DorotovicEtAl2008}
{Dorotovi{\v{c}}}, I., {Erd{\'e}lyi}, R., \& {Karlovsk{\'y}}, V. 2008, in IAU
  Symposium, Vol. 247, Waves \& Oscillations in the Solar Atmosphere: Heating
  and Magneto-Seismology, ed. R.~{Erd{\'e}lyi} \& C.~A. {Mendoza-Briceno},
  351--354

\bibitem[{{Edwin} \& {Roberts}(1983)}]{Edwin&Roberts1983}
{Edwin}, P.~M. \& {Roberts}, B. 1983, \solphys, 88, 179

\bibitem[{{Erdelyi}(1997)}]{Erdelyi1997}
{Erdelyi}, R. 1997, \solphys, 171, 49

\bibitem[{{Erd{\'e}lyi} {et~al.}(2001){Erd{\'e}lyi}, {Ballai}, \&
  {Goossens}}]{ErdelyiEtAl2001}
{Erd{\'e}lyi}, R., {Ballai}, I., \& {Goossens}, M. 2001, \aap, 368, 662

\bibitem[{{Foullon} {et~al.}(2011){Foullon}, {Verwichte}, {Nakariakov},
  {Nykyri}, \& {Farrugia}}]{FoullonEtAl2011}
{Foullon}, C., {Verwichte}, E., {Nakariakov}, V.~M., {Nykyri}, K., \&
  {Farrugia}, C.~J. 2011, \apjl, 729, L8

\bibitem[{{Freij} {et~al.}(2016){Freij}, {Dorotovi{\v{c}}}, {Morton},
  {Ruderman}, {Karlovsk{\'y}}, \& {Erd{\'e}lyi}}]{FreijEtAl2016}
{Freij}, N., {Dorotovi{\v{c}}}, I., {Morton}, R.~J., {et~al.} 2016, \apj, 817,
  44

\bibitem[{{Geeraerts} {et~al.}(2020){Geeraerts}, {Van Doorsselaere}, {Chen}, \&
  {Li}}]{GeeraertsEtAl2020}
{Geeraerts}, M., {Van Doorsselaere}, T., {Chen}, S.-X., \& {Li}, B. 2020, \apj,
  897, 120

\bibitem[{{Goossens} {et~al.}(2002){Goossens}, {Andries}, \&
  {Aschwanden}}]{GoossensEtAl2002a}
{Goossens}, M., {Andries}, J., \& {Aschwanden}, M.~J. 2002, \aap, 394, L39

\bibitem[{{Goossens} {et~al.}(2021){Goossens}, {Chen}, {Geeraerts}, {Li}, \&
  {Van Doorsselaere}}]{GoossensEtAl2021}
{Goossens}, M., {Chen}, S.~X., {Geeraerts}, M., {Li}, B., \& {Van
  Doorsselaere}, T. 2021, \aap, 646, A86

\bibitem[{{Goossens} {et~al.}(1992){Goossens}, {Hollweg}, \&
  {Sakurai}}]{GoossensEtAl1992}
{Goossens}, M., {Hollweg}, J.~V., \& {Sakurai}, T. 1992, \solphys, 138, 233

\bibitem[{{Grant} {et~al.}(2015){Grant}, {Jess}, {Moreels}, {Morton},
  {Christian}, {Giagkiozis}, {Verth}, {Fedun}, {Keys}, {Van Doorsselaere}, \&
  {Erd{\'e}lyi}}]{GrantEtAl2015}
{Grant}, S.~D.~T., {Jess}, D.~B., {Moreels}, M.~G., {et~al.} 2015, \apj, 806,
  132

\bibitem[{{Guo} {et~al.}(2019){Guo}, {Van Doorsselaere}, {Karampelas}, {Li},
  {Antolin}, \& {De Moortel}}]{GuoEtAl2019}
{Guo}, M., {Van Doorsselaere}, T., {Karampelas}, K., {et~al.} 2019, \apj, 870,
  55

\bibitem[{{Heyvaerts} \& {Priest}(1983)}]{Heyvaerts&Priest1983}
{Heyvaerts}, J. \& {Priest}, E.~R. 1983, \aap, 117, 220

\bibitem[{{Hillier} {et~al.}(2019){Hillier}, {Barker}, {Arregui}, \&
  {Latter}}]{HillierEtAl2019}
{Hillier}, A., {Barker}, A., {Arregui}, I., \& {Latter}, H. 2019, \mnras, 482,
  1143

\bibitem[{{Hillier} {et~al.}(2020){Hillier}, {Van Doorsselaere}, \&
  {Karampelas}}]{HillierEtAl2020}
{Hillier}, A., {Van Doorsselaere}, T., \& {Karampelas}, K. 2020, \apjl, 897,
  L13

\bibitem[{{Hollweg} {et~al.}(2013){Hollweg}, {Kaghashvili}, \&
  {Chandran}}]{HollwegEtAl2013}
{Hollweg}, J.~V., {Kaghashvili}, E.~K., \& {Chandran}, B. D.~G. 2013, \apj,
  769, 142

\bibitem[{{Hollweg} \& {Yang}(1988)}]{Hollweg&Yang1988}
{Hollweg}, J.~V. \& {Yang}, G. 1988, \jgr, 93, 5423

\bibitem[{{Hollweg} {et~al.}(1990){Hollweg}, {Yang}, {Cadez}, \&
  {Gakovic}}]{HollwegEtAl1990}
{Hollweg}, J.~V., {Yang}, G., {Cadez}, V.~M., \& {Gakovic}, B. 1990, \apj, 349,
  335

\bibitem[{{Karampelas} {et~al.}(2017){Karampelas}, {Van Doorsselaere}, \&
  {Antolin}}]{KarampelasEtAl2017}
{Karampelas}, K., {Van Doorsselaere}, T., \& {Antolin}, P. 2017, \aap, 604,
  A130

\bibitem[{{Karampelas} {et~al.}(2019){Karampelas}, {Van Doorsselaere},
  {Pascoe}, {Guo}, \& {Antolin}}]{KarampelasEtAl2019}
{Karampelas}, K., {Van Doorsselaere}, T., {Pascoe}, D.~J., {Guo}, M., \&
  {Antolin}, P. 2019, Frontiers in Astronomy and Space Sciences, 6, 38

\bibitem[{{Karpen} {et~al.}(1994){Karpen}, {Dahlburg}, \&
  {Davila}}]{KarpenEtAl1994}
{Karpen}, J.~T., {Dahlburg}, R.~B., \& {Davila}, J.~M. 1994, \apj, 421, 372

\bibitem[{{Kelly}(1965)}]{Kelly1965}
{Kelly}, R.~E. 1965, Journal of Fluid Mechanics, 22, 547

\bibitem[{Keys {et~al.}(2018)Keys, Morton, Jess, Verth, Grant, Mathioudakis,
  Mackay, Doyle, Christian, Keenan, \& Erd{\'{e}}lyi}]{KeysEtAl2018}
Keys, P.~H., Morton, R.~J., Jess, D.~B., {et~al.} 2018, The Astrophysical
  Journal, 857, 28

\bibitem[{{Kovitya} \& {Cram}(1983)}]{Kovitya&Cram1983}
{Kovitya}, P. \& {Cram}, L. 1983, \solphys, 84, 45

\bibitem[{{Kumar} {et~al.}(2013){Kumar}, {Innes}, \&
  {Inhester}}]{KumarEtAl2013}
{Kumar}, P., {Innes}, D.~E., \& {Inhester}, B. 2013, \apjl, 779, L7

\bibitem[{{Magyar} {et~al.}(2015){Magyar}, {Van Doorsselaere}, \&
  {Marcu}}]{MagyarEtAl2015}
{Magyar}, N., {Van Doorsselaere}, T., \& {Marcu}, A. 2015, \aap, 582, A117

\bibitem[{{Mandal} {et~al.}(2016){Mandal}, {Magyar}, {Yuan}, {Van
  Doorsselaere}, \& {Banerjee}}]{MandalEtAl2016}
{Mandal}, S., {Magyar}, N., {Yuan}, D., {Van Doorsselaere}, T., \& {Banerjee},
  D. 2016, \apj, 820, 13

\bibitem[{McLachlan(1947)}]{McLachlan1947}
McLachlan, N. 1947, Theory and Application of Mathieu Functions (Clarendon
  Press)

\bibitem[{{Moreels} {et~al.}(2015){Moreels}, {Freij}, {Erd{\'e}lyi}, {Van
  Doorsselaere}, \& {Verth}}]{MoreelsEtAl2015}
{Moreels}, M.~G., {Freij}, N., {Erd{\'e}lyi}, R., {Van Doorsselaere}, T., \&
  {Verth}, G. 2015, \aap, 579, A73

\bibitem[{{Moreels} {et~al.}(2013){Moreels}, {Goossens}, \& {Van
  Doorsselaere}}]{MoreelsEtAl2013}
{Moreels}, M.~G., {Goossens}, M., \& {Van Doorsselaere}, T. 2013, \aap, 555,
  A75

\bibitem[{{Morton} {et~al.}(2011){Morton}, {Erd{\'e}lyi}, {Jess}, \&
  {Mathioudakis}}]{MortonEtAl2011}
{Morton}, R.~J., {Erd{\'e}lyi}, R., {Jess}, D.~B., \& {Mathioudakis}, M. 2011,
  \apjl, 729, L18

\bibitem[{{Nakariakov} {et~al.}(2000){Nakariakov}, {Verwichte}, {Berghmans}, \&
  {Robbrecht}}]{NakariakovEtAl2000}
{Nakariakov}, V.~M., {Verwichte}, E., {Berghmans}, D., \& {Robbrecht}, E. 2000,
  \aap, 362, 1151

\bibitem[{{Nightingale} {et~al.}(1999){Nightingale}, {Aschwanden}, \&
  {Hurlburt}}]{NightingaleEtAl1999}
{Nightingale}, R.~W., {Aschwanden}, M.~J., \& {Hurlburt}, N.~E. 1999, \solphys,
  190, 249

\bibitem[{{Ofman} {et~al.}(1994){Ofman}, {Davila}, \&
  {Steinolfson}}]{OfmanEtAl1994}
{Ofman}, L., {Davila}, J.~M., \& {Steinolfson}, R.~S. 1994, \grl, 21, 2259

\bibitem[{{Ofman} \& {Thompson}(2011)}]{Ofman&Thompson2011}
{Ofman}, L. \& {Thompson}, B.~J. 2011, \apjl, 734, L11

\bibitem[{{Parnell} \& {De Moortel}(2012)}]{Parnell&DeMoortel2012}
{Parnell}, C.~E. \& {De Moortel}, I. 2012, Philosophical Transactions of the
  Royal Society of London Series A, 370, 3217

\bibitem[{{Pascoe} {et~al.}(2020){Pascoe}, {Goddard}, \& {Van
  Doorsselaere}}]{PascoeEtAl2020}
{Pascoe}, D.~J., {Goddard}, C.~R., \& {Van Doorsselaere}, T. 2020, Frontiers in
  Astronomy and Space Sciences, 7, 61

\bibitem[{{Pascoe} {et~al.}(2012){Pascoe}, {Hood}, {de Moortel}, \&
  {Wright}}]{PascoeEtAl2012}
{Pascoe}, D.~J., {Hood}, A.~W., {de Moortel}, I., \& {Wright}, A.~N. 2012,
  \aap, 539, A37

\bibitem[{{Pascoe} {et~al.}(2010){Pascoe}, {Wright}, \& {De
  Moortel}}]{PascoeEtAl2010}
{Pascoe}, D.~J., {Wright}, A.~N., \& {De Moortel}, I. 2010, \apj, 711, 990

\bibitem[{Priest(2014)}]{Priest2014}
Priest, E. 2014, Magnetohydrodynamics of the Sun (Cambridge University Press)

\bibitem[{{Roberts}(1973)}]{Roberts1973}
{Roberts}, B. 1973, Journal of Fluid Mechanics, 59, 65

\bibitem[{{Sakurai} {et~al.}(1991){Sakurai}, {Goossens}, \&
  {Hollweg}}]{SakuraiEtAl1991}
{Sakurai}, T., {Goossens}, M., \& {Hollweg}, J.~V. 1991, \solphys, 133, 227

\bibitem[{{Samanta} {et~al.}(2019){Samanta}, {Tian}, \&
  {Nakariakov}}]{SamantaEtAl2019}
{Samanta}, T., {Tian}, H., \& {Nakariakov}, V.~M. 2019, \prl, 123, 035102

\bibitem[{{Shi} {et~al.}(2021){Shi}, {Van Doorsselaere}, {Guo}, {Karampelas},
  {Li}, \& {Antolin}}]{ShiEtAl2021}
{Shi}, M., {Van Doorsselaere}, T., {Guo}, M., {et~al.} 2021, arXiv e-prints,
  arXiv:2101.01019

\bibitem[{Simon(2005)}]{Simon2005}
Simon, B. 2005, Trace Ideals and Their Applications, Mathematical surveys and
  monographs (American Mathematical Society)

\bibitem[{{Soler} {et~al.}(2013){Soler}, {Goossens}, {Terradas}, \&
  {Oliver}}]{SolerEtAl2013}
{Soler}, R., {Goossens}, M., {Terradas}, J., \& {Oliver}, R. 2013, \apj, 777,
  158

\bibitem[{{Soler} {et~al.}(2009){Soler}, {Oliver}, {Ballester}, \&
  {Goossens}}]{SolerEtAl2009c}
{Soler}, R., {Oliver}, R., {Ballester}, J.~L., \& {Goossens}, M. 2009, \apjl,
  695, L166

\bibitem[{Sträng(2005)}]{Strang2005}
Sträng, J.-E. 2005

\bibitem[{{Terradas} {et~al.}(2008){Terradas}, {Andries}, {Goossens},
  {Arregui}, {Oliver}, \& {Ballester}}]{TerradasEtAl2008}
{Terradas}, J., {Andries}, J., {Goossens}, M., {et~al.} 2008, \apjl, 687, L115

\bibitem[{{Van Doorsselaere} {et~al.}(2020){Van Doorsselaere}, {Srivastava},
  {Antolin}, {Magyar}, {Vasheghani Farahani}, {Tian}, {Kolotkov}, {Ofman},
  {Guo}, {Arregui}, {De Moortel}, \& {Pascoe}}]{VanDoorsselaereEtAl2020}
{Van Doorsselaere}, T., {Srivastava}, A.~K., {Antolin}, P., {et~al.} 2020,
  \ssr, 216, 140

\bibitem[{{Wang}(2011)}]{Wang2011}
{Wang}, T. 2011, \ssr, 158, 397

\bibitem[{{Wang} {et~al.}(2007){Wang}, {Innes}, \& {Qiu}}]{WangEtAl2007}
{Wang}, T., {Innes}, D.~E., \& {Qiu}, J. 2007, \apj, 656, 598

\bibitem[{{Wang} {et~al.}(2002){Wang}, {Solanki}, {Curdt}, {Innes}, \&
  {Dammasch}}]{WangEtAl2002}
{Wang}, T., {Solanki}, S.~K., {Curdt}, W., {Innes}, D.~E., \& {Dammasch}, I.~E.
  2002, \apjl, 574, L101

\bibitem[{{Wang} {et~al.}(2003{\natexlab{a}}){Wang}, {Solanki}, {Curdt},
  {Innes}, {Dammasch}, \& {Kliem}}]{WangEtAl2003a}
{Wang}, T.~J., {Solanki}, S.~K., {Curdt}, W., {et~al.} 2003{\natexlab{a}},
  \aap, 406, 1105

\bibitem[{{Wang} {et~al.}(2003{\natexlab{b}}){Wang}, {Solanki}, {Innes},
  {Curdt}, \& {Marsch}}]{WangEtAl2003b}
{Wang}, T.~J., {Solanki}, S.~K., {Innes}, D.~E., {Curdt}, W., \& {Marsch}, E.
  2003{\natexlab{b}}, \aap, 402, L17

\bibitem[{{Yu} {et~al.}(2017){Yu}, {Van Doorsselaere}, \&
  {Goossens}}]{YuEtAl2017b}
{Yu}, D.~J., {Van Doorsselaere}, T., \& {Goossens}, M. 2017, \aap, 602, A108

\bibitem[{{Zaqarashvili} {et~al.}(2015){Zaqarashvili}, {Zhelyazkov}, \&
  {Ofman}}]{ZaqarashviliEtAl2015}
{Zaqarashvili}, T.~V., {Zhelyazkov}, I., \& {Ofman}, L. 2015, \apj, 813, 123

\end{thebibliography}

\appendix

\section{Deriving the governing equation for $\c$} \label{A1}
In this appendix, we derive Eq. \eqref{Eqdivxi} from Eqs. \eqref{eq1}-\eqref{eq4}. Equations \eqref{eq1}, \eqref{eq3} and \eqref{eq4} can be rewritten using $\pmb{v} = \frac{D \pmb{\xi}}{Dt}$ to yield the following expressions for $\rho_1$, $\pmb{B}_1$ and $p_1$:

\begin{align}
\rho_1 &= -\rho_0 \c \label{rho_1}\\
\pmb{B}_1 &= -\pmb{B}_0 \l( \c \r) + \l( \pmb{B}_0 \cdot \nabla \r) \pmb{\xi} \label{B_1}\\
p_1 &= -\rho_0 v_s^2 \c \text{.} \label{p_1}
\end{align}

Using Eq. \eqref{rho_1} and Eq. \eqref{p_1}, Eq. \eqref{eq2} can be rewritten as follows:

\begin{align}
&\o_0 V_0 \sin \l( \o_0 t \r) \l( \c \r) \pmb{1}_z + \d\frac{D^2 \pmb{\xi}}{Dt^2} = \notag\\
 & \hspace{3cm} \frac{-1}{\rho_0} \nabla P_1 + i k_z v_A^2 \l[- \pmb{1}_z \l( \c \r) + i k_z \pmb{\xi} \r] \text{,} \label{6bis}
\end{align}
where $v_A = B_{0z} / \sqrt{\mu_0 \rho_0}$ is the Alfv\'en speed. By now taking the divergence on both sides of Eq. \eqref{6bis} we get the following:

\begin{equation}
i k_z \o_0 V_0 \sin \l( \o_0 t \r) \l( \c \r) + \d\frac{D^2 \l( \c \r)}{Dt^2} = \frac{-1}{\rho_0} \nabla^2 P_1 \text{,} \label{7}
\end{equation}
where $\nabla^2$ denotes the Laplace operator.

Next, from the definition of total pressure $P_1 = p_1 + \pmb{B}_0 \cdot \pmb{B}_1 / \mu_0$ we obtain the following equation with the use of Eqs. \eqref{B_1} and \eqref{p_1}:

\begin{equation}
\d\frac{1}{\rho_0} P_1 + (v_A^2 + v_s^2) \l( \c \r) - i k_z v_A^2 \xi_z = 0 \text{.} \label{8}
\end{equation}

Taking the $z$-component of Eq. \eqref{6bis} and using Eq. \eqref{8} yields

\begin{equation}
\d\frac{D^2 \xi_z}{Dt^2} = - \o_0 V_0 \sin \l( \o_0 t \r) \l( \c \r) + i k_z v_s^2 \l( \c \r) \text{,} \label{9}
\end{equation}
while taking twice the Lagrangian derivative on both sides of Eq. \eqref{8} and rearanging the terms yields

\begin{equation}
\d\frac{D^2 \xi_z}{Dt^2}  = \frac{1}{i k_z v_A^2 \rho_0} \frac{D^2 P_1}{Dt^2} +\frac{v_A^2 + v_s^2}{i k_z v_A^2} \frac{D^2 \l( \c \r)}{Dt^2} \text{.} \label{10}
\end{equation}
We can then combine Eq. \eqref{9} and Eq. \eqref{10} into the following equation:

\begin{align}
&\d\frac{1}{i k_z v_A^2 \rho_0} \frac{D^2 P_1}{Dt^2} +\frac{v_A^2 + v_s^2}{i k_z v_A^2} \frac{D^2 \l( \c \r)}{Dt^2} = \notag\\
& \hspace{2.5cm} - \o_0 V_0 \sin \l( \o_0 t \r) \l( \c \r) + i k_z v_s^2 \l( \c \r) \text{.} \label{11}
\end{align}

If we now take twice the Lagrangian derivative on both sides of Eq. \eqref{7} and we take the Laplacian on both sides of Eq. \eqref{11}, we can combine the obtained equations, yielding Eq. \eqref{Eqdivxi}:

\begin{align}
&\d\frac{D^4 \l( \c \r)}{Dt^4} + i k_z \o_0 V_0 \frac{D^2 \l( \sin \l(\o_0 t \r) \l( \c \r) \r)}{Dt^2} \notag\\
& \qquad - \l( v_A^2 + v_s^2 \r) \frac{D^2}{Dt^2} \l( \frac{\pa^2 \l( \c \r)}{\pa x^2} \r) + k^2 \l( v_A^2 + v_s^2 \r) \frac{D^2 \l( \c \r)}{Dt^2} \notag\\
& \qquad - i k_z V_0 \o_0 \sin \l( \o_0 t \r) v_A^2 \frac{\pa^2 \l( \c \r)}{\pa x^2} \notag\\
& \qquad + i k_z  k^2 V_0 \o_0 \sin \l(\o_0 t \r)  v_A^2 \l( \c \r) - k_z^2 v_A^2 v_s^2 \frac{\pa^2 \l( \c \r)}{\pa x^2} \notag \\
& \qquad + k_z^2 k^2 v_A^2 v_s^2 \l( \c \r) = 0 \text{,}
\end{align}
where $k = \sqrt{k_y^2 + k_z^2}$.

\section{Deriving the expression for $\Delta$} \label{ExtPow}

In this appendix, we show formula \eqref{DeltaFormula} for $\Delta$. We recall the definition of $\Delta$, Eq. \eqref{Delta}, for the reader's convenience:

\begin{equation} \label{Delta2}
\Delta = \begin{vmatrix}
\ddots & \ddots & &  &  0\\ \ddots & 1 & \varepsilon_{-1} &  & \\  & -\varepsilon_0 & 1 & \varepsilon_0 & \\ &  & -\varepsilon_1 & 1 & \ddots \\ 0 &  &  & \ddots & \ddots
\end{vmatrix} \text{.}
\end{equation}
Denoting with $A$ the operator defined by the infinite matrix corresponding to the determinant in the right-hand side of Eq. \eqref{Delta2}, we explained in Section \ref{Gfunction} that $A-I$ (with $I$ the identity operator) is a trace class operator on $\ell^2(\mathbb{Z})$ (except in the poles of the $\varepsilon_j$). This is the Hilbert space of square-summable sequences of complex numbers with entire index, and with inner product $\langle \cdot, \cdot \rangle$ defined by

\begin{equation}
\langle x, y \rangle = \d\sum_{l=-\infty}^{\infty} x(l) \overline{y(l)}
\end{equation}
for $x,y \in \ell^2(\mathbb{Z})$. To work out the right-hand side of Eq. \eqref{Delta2}, we use a result from Fredholm theory. It states that, for a trace class operator $M$, the following holds:

\begin{equation} \label{FredholmDet}
\det(I+M) \; = \; \d\sum_{n=0}^{\infty} \text{Tr} \l( \Lambda^n \l( M \r) \r) \text{,}
\end{equation}
where $\text{Tr} \l( \Lambda^n \l( M \r) \r)$ is the trace of the $n$th exterior power of $M$ \citep{Simon2005}.

The $n$th exterior power $\Lambda^n (\ell^2(\mathbb{Z}))$ (with $n \in \mathbb{N}_0$) of $\ell^2(\mathbb{Z})$ is the Hilbert space whose elements are of the form $v_1 \wedge ... \wedge v_n$, that is to say, the exterior product of $v_1$, ..., $v_n$ (in that order) where $v_i \in \ell^2(\mathbb{Z})$ for every $i \in \{1, ..., n \}$, and with a uniquely defined inner product $\langle \cdot , \cdot \rangle^{\otimes n}$ that is determined by the inner product on $\ell^2(\mathbb{Z})$. The vector space $\Lambda^n (\ell^2(\mathbb{Z}))$ is the subspace of the tensor product $\ell^2(\mathbb{Z}) \otimes ... \otimes \ell^2(\mathbb{Z})$ (n times) whose elements are totally antisymmetric tensors of rank $n$. Furthermore, the 0th exterior power $\Lambda^0 (\ell^2(\mathbb{Z}))$ of $\ell^2(\mathbb{Z})$ is $\mathbb{C}$.

For a trace class operator $M$ on $\ell^2(\mathbb{Z})$, the $n$th exterior power of $M$ is defined by

\begin{align}
\Lambda^n (M): \; &\Lambda^n ( \ell^2 (\mathbb{Z})) \to \Lambda^n ( \ell^2 (\mathbb{Z})): \notag\\
&\hspace{2cm} v_1 \wedge ... \wedge v_n \mapsto M v_1 \wedge ... \wedge M v_n  \label{Lambda^n(M)}
\end{align}
and is a trace class operator on $\Lambda^n (\ell^2(\mathbb{Z}))$. Now, if $\{e_j \}_{j \in \mathbb{Z}}$ is an orthonormal basis for a Hilbert space $H$, then the trace of a trace class operator $M$ on $H$ is defined as \citep{Conway1990}:

\begin{equation} \label{trace}
\text{Tr}(M) = \d\sum_{j=-\infty}^{\infty} \langle M e_j, e_j \rangle \text{.}
\end{equation}
We also have that, if $\{ e_j \}_{j \in \mathbb{Z}}$ is an orthonormal basis of $\ell^2(\mathbb{Z})$, then $\{e_{j_1} \wedge ... \wedge e_{j_n} \}_{j_1 < \cdots < j_n}$ is an orthonormal basis of $\Lambda^n ( \ell^2 (\mathbb{Z}))$ and

\begin{equation} \label{innprodextprod}
\d \langle e_{j_1} \wedge ... \wedge e_{j_n}, f_{1} \wedge ... \wedge f_{n} \rangle^{\otimes n} = \det \l(\langle e_{j_k}, f_m \rangle_{1 \leq k,m \leq n} \r) \text{,}
\end{equation}
for any $f_{1} \wedge ... \wedge f_{n} \in \Lambda^n ( \ell^2 (\mathbb{Z}))$ \citep{Simon2005}. Hence, from Eqs. \eqref{Lambda^n(M)}, \eqref{trace} and \eqref{innprodextprod}, we find for the trace class operator $\Lambda^n \l( M \r)$ on $\Lambda^n (\ell^2(\mathbb{Z}))$ that

\begin{align}
&\text{Tr} \l(\Lambda^n \l( M \r) \r) \notag\\
&= \d\sum_{j_1 = -\infty}^{\infty} \sum_{j_2 = j_1 + 1}^{\infty} ... \sum_{j_n = j_{n-1} + 1}^{\infty} \ip{ \Lambda^n \l( M \r) \l( e_{j_1} \wedge ... \wedge e_{j_n} \r) , e_{j_1} \wedge ... \wedge e_{j_n} }^{\otimes n} \notag \\
&= \d\sum_{j_1 = -\infty}^{\infty} \sum_{j_2 = j_1 + 1}^{\infty} ... \sum_{j_n = j_{n-1} + 1}^{\infty} \ip{ M e_{j_1} \wedge ... \wedge M e_{j_n} ,  e_{j_1} \wedge ... \wedge e_{j_n} }^{\otimes n} \notag \\
&= \d\sum_{j_1 = -\infty}^{\infty} \sum_{j_2 = j_1 + 1}^{\infty} ... \sum_{j_n = j_{n-1} + 1}^{\infty} \begin{vmatrix} a_{1,1} & \cdots & a_{1,n} \\ \vdots & & \vdots\\  a_{n,1} & \cdots & a_{n,n} \end{vmatrix} \text{,} \label{TrExpr1}
\end{align}
where $a_{k,m} = \langle M e_{j_k}, e_{j_m} \rangle$ for $k,m \in \mathbb{N}_0$.

Looking at Eq. \eqref{Delta2}, we see that the trace class operator $M=A-I$ is defined by
 
\begin{equation}
M : \ell^2\l(\mathbb{Z} \r) \to \ell^2\l(\mathbb{Z} \r) : \begin{pmatrix} \vdots \\ \varphi(l) \\ \vdots  \end{pmatrix} \mapsto \begin{pmatrix} \vdots \\ \varepsilon_l \l[ \varphi \l( l+1 \r) - \varphi \l( l-1 \r) \r] \\ \vdots  \end{pmatrix} \text{.}
\end{equation}
For the orthonormal basis $\{ e_j \}_{j \in \mathbb{Z}}$ of $\ell^2(\mathbb{Z})$ defined by $e_j(l) = \delta_{jl}$ for all $j,l \in \mathbb{Z}$, we then have the following:

\begin{equation}
M e_{j_k} = M \begin{pmatrix} \vdots \\ 0 \\ 0 \\ 1 \\ 0 \\ 0 \\ \vdots  \end{pmatrix} = \begin{pmatrix} \vdots \\ 0 \\ \varepsilon_{j_k-1} \\ 0 \\ -\varepsilon_{j_k+1} \\ 0 \\ \vdots  \end{pmatrix} \text{,}
\end{equation}
for every $k \in \mathbb{N}_0$. This means that

\begin{align}
\ip{M e_{j_k},e_{j_m}} &= \d\sum_{l=-\infty}^{\infty} \l(M e_{j_k} \r)(l) \; \overline{e_{j_m}(l)} \notag \\
&= \begin{cases} \varepsilon_{j_k-1} & \text{if } j_m = j_k -1 \\ - \varepsilon_{j_k+1} & \text{if } j_m = j_k+1 \\ 0 & \text{else}\end{cases} \text{.} \label{innprod}
\end{align}
We thus find that $\langle M e_{j_k}, e_{j_m} \rangle = 0$ if $\abs{k-m} \ge 2$ (since $j_1 < ... < j_n$) or $k=m$. Hence, from Eq. \eqref{TrExpr1},

\begin{equation} \label{traceExpr2}
\text{Tr} \l(\Lambda^n \l( M \r) \r) =  \d\sum_{j_1 = -\infty}^{\infty} \sum_{j_2 = j_1 + 1}^{\infty} \ldots \sum_{j_n = j_{n-1} + 1}^{\infty} \Delta_{j_1,...,j_n} \text{,}
\end{equation}
where we define for any $q \in \mathbb{N}_0$ and $s_1,..., s_q \in \mathbb{N}_0$

\begin{equation} \label{Delta_js}
\Delta_{j_{s_1},...,j_{s_q}} = \begin{tikzpicture}[baseline=(current bounding box.center)]
\matrix (m) [matrix of math nodes,nodes in empty cells,right delimiter={|},left delimiter={|} ]{
0                               & a_{s_1,s_2} & 0                                     &                                         & 0                                     \\
a_{s_2,s_1} &                                  &                                        &                                         &                                        \\
 0                              &                                  &                                        &                                         & 0                                      \\
                                 &                                  &                                        &                                         &a_{s_{q-1},s_q} \\
0                               &                                  & 0                                     & a_{s_q,s_{q-1}}  &  0                                    \\
} ;
\draw[loosely dotted][thick] (m-1-1)-- (m-5-5);
\draw[loosely dotted][thick] (m-1-2)-- (m-4-5);
\draw[loosely dotted][thick] (m-2-1)-- (m-5-4);
\draw[loosely dotted][thick] (m-3-1)-- (m-5-1);
\draw[loosely dotted][thick] (m-5-1)-- (m-5-3);
\draw[loosely dotted][thick] (m-3-1)-- (m-5-3);
\draw[loosely dotted][thick] (m-1-3)-- (m-1-5);
\draw[loosely dotted][thick] (m-1-5)-- (m-3-5);
\draw[loosely dotted][thick] (m-1-3)-- (m-3-5);
\end{tikzpicture} \text{.}
\end{equation}

We can now rewrite the determinants $\Delta_{j_1,...,j_n}$ composing the terms of the sum in Eq. \eqref{traceExpr2}:

\begin{eqnarray}
\Delta_{j_1,...,j_n} &=&\begin{tikzpicture}[baseline=(current bounding box.center)]
\matrix (m) [matrix of math nodes,nodes in empty cells,right delimiter={|},left delimiter={|} ]{
0                               & a_{1,2} & 0                                     &                                         & 0                                     \\
a_{2,1} &                                  &                                        &                                         &                                        \\
 0                              &                                  &                                        &                                         & 0                                      \\
                                 &                                  &                                        &                                         &a_{n-1,n} \\
0                               &                                  & 0                                     & a_{n,n-1}  &  0                                    \\
} ;
\draw[loosely dotted][thick] (m-1-1)-- (m-5-5);
\draw[loosely dotted][thick] (m-1-2)-- (m-4-5);
\draw[loosely dotted][thick] (m-2-1)-- (m-5-4);
\draw[loosely dotted][thick] (m-3-1)-- (m-5-1);
\draw[loosely dotted][thick] (m-5-1)-- (m-5-3);
\draw[loosely dotted][thick] (m-3-1)-- (m-5-3);
\draw[loosely dotted][thick] (m-1-3)-- (m-1-5);
\draw[loosely dotted][thick] (m-1-5)-- (m-3-5);
\draw[loosely dotted][thick] (m-1-3)-- (m-3-5);
\end{tikzpicture} \label{TrPowerprev} \\
&=& - a_{1,2} \begin{tikzpicture}[baseline=(current bounding box.center)]
\matrix (m) [matrix of math nodes,nodes in empty cells,right delimiter={|},left delimiter={|} ]{
a_{2,1}    & a_{2,3}          & 0                       &                      &                                         & 0                                     \\
0            & 0                   & a_{3,4}               & 0                  &                                         & 0                                     \\
              & a_{4,3}          &                          &                      &                                         &                                        \\
              & 0                   &                          &                      &                                         & 0                                      \\
              &                      &                          &                      &                                         &a_{n-1,n} \\
0            & 0                   &                          & 0                   & a_{n,n-1}                     &  0                                    \\
} ;
\draw[loosely dotted][thick] (m-2-2)-- (m-6-6);
\draw[loosely dotted][thick] (m-2-3)-- (m-5-6);
\draw[loosely dotted][thick] (m-3-2)-- (m-6-5);
\draw[loosely dotted][thick] (m-4-2)-- (m-6-2);
\draw[loosely dotted][thick] (m-6-2)-- (m-6-4);
\draw[loosely dotted][thick] (m-4-2)-- (m-6-4);
\draw[loosely dotted][thick] (m-2-4)-- (m-2-6);
\draw[loosely dotted][thick] (m-2-6)-- (m-4-6);
\draw[loosely dotted][thick] (m-2-4)-- (m-4-6);
\draw[loosely dotted][thick] (m-2-1)-- (m-6-1);
\draw[loosely dotted][thick] (m-1-3)-- (m-1-6);
\end{tikzpicture} \notag \\
&=& - a_{1,2} a_{2,1}\begin{tikzpicture}[baseline=(current bounding box.center)]
\matrix (m) [matrix of math nodes,nodes in empty cells,right delimiter={|},left delimiter={|} ]{
0                               & a_{3,4} & 0                                     &                                         & 0                                     \\
a_{4,3} &                                  &                                        &                                         &                                        \\
 0                              &                                  &                                        &                                         & 0                                      \\
                                 &                                  &                                        &                                         &a_{n-1,n} \\
0                               &                                  & 0                                     & a_{n,n-1}  &  0                                    \\
} ;
\draw[loosely dotted][thick] (m-1-1)-- (m-5-5);
\draw[loosely dotted][thick] (m-1-2)-- (m-4-5);
\draw[loosely dotted][thick] (m-2-1)-- (m-5-4);
\draw[loosely dotted][thick] (m-3-1)-- (m-5-1);
\draw[loosely dotted][thick] (m-5-1)-- (m-5-3);
\draw[loosely dotted][thick] (m-3-1)-- (m-5-3);
\draw[loosely dotted][thick] (m-1-3)-- (m-1-5);
\draw[loosely dotted][thick] (m-1-5)-- (m-3-5);
\draw[loosely dotted][thick] (m-1-3)-- (m-3-5);
\end{tikzpicture} \notag \\
&=& - a_{1,2} a_{2,1} \Delta_{j_3,...,j_n} \label{TrPower} \text{,}
\end{eqnarray}
where we expanded along the first row to find the second equality and along the first column to find the third equality. Clearly, we can omit terms that are $0$ from the sum in the right-hand side of Eq. \eqref{traceExpr2}. Now, the factors $a_{1,2}$ and $a_{2,1}$ are different from $0$ only if $j_2 = j_1+1$, in which case 

\begin{align*}
&a_{1,2} = \ip{ M e_{j_1}, e_{j_1+1}} = - \varepsilon_{j_1 + 1} \text{, and}\\
& a_{2,1} = \ip{ M e_{j_1+1}, e_{j_1}} = \varepsilon_{j_1} \text{.}
\end{align*}
This means that, from Eqs. \eqref{traceExpr2}, \eqref{Delta_js} and \eqref{TrPower}, we have

\begin{equation} \label{traceExpr3}
\text{Tr} \l(\Lambda^n \l( M \r) \r) = \d\sum_{j_1 = -\infty}^{\infty} \varepsilon_{j_1} \varepsilon_{j_1 + 1} \sum_{j_3 = j_1 + 2}^{\infty} \sum_{j_4 = j_3 + 1}^{\infty} \ldots \sum_{j_n = j_{n-1} + 1}^{\infty} \Delta_{j_3,...,j_n} \text{.}
\end{equation}
We can now make a distinction between even and odd $n$. Indeed, if $n = 2p$ for a $p \in \mathbb{N}_0$, we find the following by continuing to expand every determinant of the form \eqref{Delta_js} in the sum on the right-hand side of Eq. \eqref{traceExpr3} in the same way as we went from Eq. \eqref{TrPowerprev} to \eqref{TrPower}:

\begin{equation} \label{TraceEven}
\text{Tr} \l(\Lambda^{2p} \l( M \r) \r) = \d\sum_{j_1 = - \infty}^{\infty} \d\sum_{j_2 = j_1 + 2}^{\infty} \ldots \d\sum_{j_p = j_{p-1} + 2}^{\infty} \l( \d\prod_{l=1}^{p}  \varepsilon_{j_l} \varepsilon_{j_l +1} \r) \text{.}
\end{equation}
In contrast to this, if $n=2p+1$ for a $p \in \mathbb{N}$, we find

\begin{align}
\text{Tr} \l(\Lambda^{2p+1} \l( M \r) \r) &= \d\sum_{j_1 = - \infty}^{\infty} \d\sum_{j_2 = j_1 + 2}^{\infty} \ldots \d\sum_{j_p = j_{p-1} + 2}^{\infty} \l( \d\prod_{l=1}^{p}  \varepsilon_{j_l} \varepsilon_{j_l +1} \r) 0 \notag\\
&= 0 \text{.} \label{TrUneven}
\end{align}

Hence, from Eqs. \eqref{FredholmDet}, \eqref{TraceEven} and \eqref{TrUneven}, and from the fact that $\text{Tr} \l(\Lambda^0 \l( M \r) \r) = 1$ \citep{Simon2005}, we now find Eq. \eqref{DeltaFormula}:

\begin{equation}
\Delta \; = \; 1 + \d\sum_{n=1}^{\infty} \l[ \d\sum_{j_1 = - \infty}^{\infty} \d\sum_{j_2 = j_1 + 2}^{\infty} \ldots \d\sum_{j_n = j_{n-1} + 2}^{\infty} \l( \d\prod_{l=1}^n  \varepsilon_{j_l} \varepsilon_{j_l +1} \r) \r] \text{.}
\end{equation}

\section{Deriving solutions for $G_1$, $G_2$ and $G_3$} \label{AppendixG1G2G3}

In this appendix, we show how to derive the solutions to the temporal functions $G_1$, $G_2$ and $G_3$. We recall the equations of interest to this matter, Eqs. \eqref{eq1}-\eqref{eq3}, Eqs. \eqref{xix}-\eqref{xiz} and Eq. \eqref{P1} for the reader's convenience:

\begin{align}
&\d\frac{D \rho_1}{D t} +\rho_0 \l( \nabla \cdot \pmb{v}_1 \r) = 0\text{,} \label{B1}\\
& \rho_1 \frac{\partial \pmb{v}_0}{\partial t}  + \rho_0 \frac{D \pmb{v}_1}{D t} = -\nabla P_1 + \frac{1}{\mu_0} \left( \pmb{B}_0 \cdot \nabla \right) \pmb{B}_1 \text{,} \label{B2}\\
&\frac{D \pmb{B}_1}{D t} = - \pmb{B}_0 \left( \nabla \cdot \pmb{v}_1 \right) + \left( \pmb{B}_0 \cdot \nabla \right) \pmb{v}_1 \text{,} \label{B3}\\
&\xi_x = - \d\frac{i m}{k^2 + m^2} \; \tilde{F}(x)\; h_1(t) \; \e \text{,} \label{B5}\\
&\xi_y = - \d\frac{i k_y}{k^2 + m^2} \; F(x)\; h_2(t) \; \e \text{,} \label{B6}\\
&\xi_z = - \d\frac{i k_z}{k^2 + m^2} \; F(x)\; h_3(t) \; \e \text{,} \label{B7}\\
&P_1 = \rho_0 \; \l( \d\frac{k_z^2 v_A^2}{k^2 + m^2} h_3(t) \; - \; \l( v_A^2 + v_s^2 \r) h(t) \r) \notag\\ 
& \hspace{4cm} F(x) \; \e \text{.} \label{B8}
\end{align}
We recall Eq. \eqref{6bis} from Appendix \ref{A1}:

\begin{align}
&\o_0 V_0 \sin \l( \o_0 t \r) \l( \c \r) \pmb{1}_z + \d\frac{D^2 \pmb{\xi}}{Dt^2} = \notag\\
 & \hspace{3cm} \frac{-1}{\rho_0} \nabla P_1 + i k_z v_A^2 \l[- \pmb{1}_z \l( \c \r) + i k_z \pmb{\xi} \r] \text{.} \label{B9}
\end{align}

Replacing $h_i(t)$ by their definition $G_i(t) g(t)$ for every $i \in \{1,2,3 \}$ (where $g(t) = \exp \l\{ \frac{-i k_z V_0 \sin \l( \o_0 t \r)}{\o_0} \r\}$)  and inserting Eqs. \eqref{B5}-\eqref{B8} into the $x$-component, $y$-component and $z$-component of Eq. \eqref{B9}, we find that the following equations have to hold:

\begin{align}
&\d\frac{d^2 G_1(t)}{dt^2} + K^2 \l(v_A^2 + v_s^2 \r) G(t) + k_z^2 v_A^2 \l[ G_1(t) - G_3(t) \r] = 0\text{,} \label{B.G1}\\
&\d\frac{d^2 G_2(t)}{dt^2} + K^2 \l(v_A^2 + v_s^2 \r) G(t) + k_z^2 v_A^2 \l[ G_2(t) - G_3(t) \r] = 0\text{,} \label{B.G2}\\
&\d\frac{d^2 G_3(t)}{dt^2} + K^2 \l(v_s^2 + \d\frac{i \o_0 V_0}{k_z} \sin \l( \o_0 t \r) \r) G(t) = 0 \text{,} \label{B.G3}
\end{align}
with $K = \sqrt{k^2 + m^2}$.

We see that $G_1$ and $G_2$ obey the same differential equation and that they thus have the same general solution. They depend on $G_3$, which by solving Eq. \eqref{B.G3} can be found to be equal to

\begin{equation} \label{B.G3eq}
G_3(t) = - \d\frac{K^2}{k_z} \d\iint \l(i \o_0 V_0 \sin \l( \o_0 t \r) + k_z v_s^2 \r) G(t) \; dt \; dt \text{.}
\end{equation}
We recall that we defined $\t = \o_0 t$. So, since $G(\t) = e^{\mu \t} \sum_{j = -\infty}^{\infty} \varphi_j e^{i j \t}$, we find that

\begin{align} 
G_3(\t) &= -\d\frac{K^2}{2 k_z \o_0^2} e^{\mu \t} \d\sum_{j = -\infty}^{\infty} \d\frac{V_0 \o_0 \l( \varphi_{j+1} - \varphi_{j-1} \r) - 2 k_z v_s^2 \varphi_j}{\l( j + \nu \r)^2} \; e^{i j \t} \notag \\
&= \d\frac{K^2}{k_z^2} e^{\mu \t} \d\sum_{j = -\infty}^{\infty} \l(1 - \d\frac{\l( k_y^2+ m^2 \r) v_s^2}{\l(j + \nu \r)^2 \o_0^2 - K^2 v_A^2} \r) \; \varphi_j \; e^{i j \t} \text{,} \label{B.G3exp}
\end{align}
where we used Eq. \eqref{recursion} to find the second equality. 

Now, solving Eq. \eqref{B.G1}, we find that

\begin{align}
G_1(t) &= \d\frac{1}{k_z v_A} \Bigg\{ \cos \l( k_z v_A t \r) \notag \\
& \qquad  \l[ \d \int \sin \l( k_z v_A t \r) \l( K^2 \l( v_A^2 + v_s^2 \r) G(t)  - k_z^2 v_A^2 G_3(t) \r) dt \r] \notag\\
& \quad- \sin \l( k_z v_A t \r) \notag \\
& \qquad \l[ \d \int \cos \l( k_z v_A t \r) \l( K^2 \l( v_A^2 + v_s^2 \r) G(t) - k_z^2 v_A^2 G_3(t) \r) dt \r] \Bigg\} \text{.} \label{B.G1eq}
\end{align}
Having derived expression \eqref{B.G3exp}, we can also work out Eq. \eqref{B.G1eq} and find the following solution for $G_1$:

\begin{equation} \label{B.G1exp}
G_1(\t) = K^2 v_s^2 \; e^{\mu \t} \d\sum_{j = -\infty}^{\infty} \d\frac{1}{\l(j + \nu \r)^2 \o_0^2 - K^2 v_A^2} \; \varphi_j \; e^{i j \t} \text{.}
\end{equation}
Analogously,

\begin{equation} \label{B.G1exp}
G_2(\t) = K^2 v_s^2 \; e^{\mu \t} \d\sum_{j = -\infty}^{\infty} \d\frac{1}{\l(j + \nu \r)^2 \o_0^2 - K^2 v_A^2} \; \varphi_j \; e^{i j \t} \text{.}
\end{equation}

%\begin{thebibliography}{}
%
%  \bibitem[Baker(1966)]{baker} Baker, N. 1966,
%      in Stellar Evolution,
%      ed.\ R. F. Stein,\& A. G. W. Cameron
%      (Plenum, New York) 333
%
%   \bibitem[Balluch(1988)]{balluch} Balluch, M. 1988,
%      A\&A, 200, 58
%
%   \bibitem[Cox(1980)]{cox} Cox, J. P. 1980,
%      Theory of Stellar Pulsation
%      (Princeton University Press, Princeton) 165
%
%   \bibitem[Cox(1969)]{cox69} Cox, A. N.,\& Stewart, J. N. 1969,
%      Academia Nauk, Scientific Information 15, 1
%
%   \bibitem[Mizuno(1980)]{mizuno} Mizuno H. 1980,
%      Prog. Theor. Phys., 64, 544
%   
%   \bibitem[Tscharnuter(1987)]{tscharnuter} Tscharnuter W. M. 1987,
%      A\&A, 188, 55
%  
%   \bibitem[Terlevich(1992)]{terlevich} Terlevich, R. 1992, in ASP Conf. Ser. 31, 
%      Relationships between Active Galactic Nuclei and Starburst Galaxies, 
%      ed. A. V. Filippenko, 13
%
%   \bibitem[Yorke(1980a)]{yorke80a} Yorke, H. W. 1980a,
%      A\&A, 86, 286
%
%   \bibitem[Zheng(1997)]{zheng} Zheng, W., Davidsen, A. F., Tytler, D. \& Kriss, G. A.
%      1997, preprint
%\end{thebibliography}

\end{document}